\documentclass[conference,compsoc]{IEEEtran}
\usepackage[]{sty/wspr}
\usepackage{float}
\ifCLASSOPTIONcompsoc
\usepackage[nocompress]{cite}
\else
\usepackage{cite}
\fi
\usepackage[pdftex]{graphicx}
\usepackage{booktabs}
\usepackage{multirow}
\usepackage{array}

\usepackage{fancyhdr}

\begin{document}

\clearpage{}\newcommand{\RoBo}{Robocall Observatory}
\newcommand{\RoBos}{Robocall Observatory's}
\newcommand{\Robo}{Robocall Observatory}
\newcommand{\robo}{Robocall Observatory}
\newcommand{\roboTotalCalls}{10,056,650}
\newcommand{\roboStartTime}{Apr 1st, 2019}
\newcommand{\roboEndTime}{May 25th, 2024}
\newcommand{\roboTotalCampaigns}{XXX}

\newcommand{\roboTotalCallsSinceSS}{3,234,362}
\newcommand{\roboCallsPerDay}{5,349}
\newcommand{\roboStartSS}{1st Nov 2021}
\newcommand{\roboSSRawCalls}{179,544}
\newcommand{\roboSSCalls}{51.2\%}
\newcommand{\roboAnsweredCalls}{2,302,606}
\newcommand{\roboASS}{983,519}
\newcommand{\roboBSS}{698,278}
\newcommand{\roboCSS}{307,986}
\newcommand{\roboNoneSS}{1,244,579}
\newcommand{\roboPrecentValid}{25.5\%}
\newcommand{\whosCallingPrecentValid}{37.25\%}
\newcommand{\rraptor}{\textsc{Rraptor}}
\newcommand{\rraptors}{\textsc{Rraptor}'s}
\newcommand{\rraptorTotalCalls}{948,775}
\newcommand{\rraptorStartTime}{2024-01-01}
\newcommand{\rraptorEndTime}{2024-05-23}
\newcommand{\rraptorTotalCampaigns}{15,475}
\newcommand{\rraptorCallsPerDay}{6,682}
\newcommand{\rraptorCallsWithSubstantialAudio}{343,141}
\newcommand{\rraptorTotalCallsClustered}{107,531}
\newcommand{\rraptorPercentageOfCallsClustered}{31.33\%}
\newcommand{\rraptorMeanCampaignSize}{6.94}
\newcommand{\rraptorMaxCampaignSize}{3714}
\newcommand{\rraptorSSAttestationA}{44.3\%}
\newcommand{\rraptorSSAttestationB}{12.8\%}
\newcommand{\rraptorSSAttestationC}{24.5\%}
\newcommand{\rraptorSSAttestationNone}{18.4\%}
\newcommand{\rraptorSSAttestationACount}{420,639}
\newcommand{\rraptorSSAttestationBCount}{121,026}
\newcommand{\rraptorSSAttestationCCount}{232,344}
\newcommand{\rraptorSSAttestationNoneCount}{174,766}
\newcommand{\rraptorMaxAudioDuration}{90 seconds}
\newcommand{\rraptorMoreThanTenpercent}{633,818}
\newcommand{\rraptorMoreThanFiveseconds}{346,743}
\newcommand{\rraptorMoreThanTenpercentAndMoreThanFivesec}{344,254}
\newcommand{\rraptorMoreThanTenpercentAndMoreThanFivesecPercentage}{36.28 \%}
\newcommand{\rraptorMoreThanTenpercentPercentage}{66.8\%}
\newcommand{\rraptorMoreThanFivesecPercentage}{36.3\%}
\newcommand{\rraptorSilhouetteScore}{0.499}
\newcommand{\rraptorCalinskiScore}{540.154}
\newcommand{\rraptorEnglishcallsPercetage}{96.9\%}
\newcommand{\rraptorSpanishcallsPercentage}{1.4\%}
\newcommand{\rraptorMandarincallsPercentage}{0.9\%}
\newcommand{\rraptorEnglishcalls}{104,094}
\newcommand{\rraptorSpanishcalls}{1,454}
\newcommand{\rraptorMandarincalls}{1,004}
\newcommand{\rraptorTotalCallbackNumbersClustrered}{2,325}
\newcommand{\rraptorCampaignsWithcallbackNumbers}{2,519}
\newcommand{\rraptorCampaignsWithcallbackNumbersVocalized}{6}

\newcommand{\ftcTotalReports}{13,707,524}
\newcommand{\ftcStartTime}{2020-02-14}
\newcommand{\ftcEndTime}{2024-05-23}
\newcommand{\ftcReportsPerDay}{8,787}
\newcommand{\ppone}{\textsc{PPoNE}}
\newcommand{\pponeTotalTracebacks}{1,353}
\newcommand{\pponeStartTime}{2020-12-30}
\newcommand{\pponeEndTime}{2023-10-11}
\newcommand{\pponeTotalCases}{31}
\newcommand{\pponeReclustedTotalCampaigns}{174}
\newcommand{\pponeHumanLabels}{118}
\newcommand{\silhouetteScoreCosine}{0.67}
\newcommand{\silhouetteScoreEuclidean}{0.54}
\newcommand{\calinskiHarabaszScore}{211.71}
\newcommand{\daviesBouldinScore}{0.7}

\newcommand{\irsTotalCalls}{668,991}
\newcommand{\irsStartTime}{2012-11-07}
\newcommand{\irsEndTime}{2023-02-14}
\newcommand{\irsReportsPerDay}{178}

\newcommand{\monthlyHoneypotMean}{12.9e9}
\newcommand{\monthlyHoneypotR}{0.227}
\newcommand{\monthlyHoneypotIntercept}{284600.067}
\newcommand{\monthlyHoneypotCoeff}{-127}

\newcommand{\comprehensiveSignedMeanSqError}{0.043}
\newcommand{\comprehensiveSignedRTwo}{0.035}
\newcommand{\comprehensiveSignedIntercept}{0.558}
\newcommand{\comprehensiveSignedCoeff}{0.00015}

\newcommand{\comprehensiveUnsignedMeanSqError}{0.024}
\newcommand{\comprehensiveUnsignedRTwo}{0.042}
\newcommand{\comprehensiveUnsignedIntercept}{0.366}
\newcommand{\comprehensiveUnsignedCoeff}{-0.00012}

\newcommand{\rraptorAllSignedMeanSqError}{0.009}
\newcommand{\rraptorAllSignedRTwo}{0.022}
\newcommand{\rraptorAllSignedIntercept}{0.774}
\newcommand{\rraptorAllSignedCoeff}{0.00035}

\newcommand{\rraptorASignedMeanSqError}{0.004}
\newcommand{\rraptorASignedRTwo}{0.030}
\newcommand{\rraptorASignedIntercept}{0.424}
\newcommand{\rraptorASignedCoeff}{0.00026}

\newcommand{\rraptorBSignedMeanSqError}{0.001}
\newcommand{\rraptorBSignedRTwo}{0.004}
\newcommand{\rraptorBSignedIntercept}{0.119}
\newcommand{\rraptorBSignedCoeff}{0.00004}

\newcommand{\rraptorCSignedMeanSqError}{0.002}
\newcommand{\rraptorCSignedRTwo}{0.002}
\newcommand{\rraptorCSignedIntercept}{0.231}
\newcommand{\rraptorCSignedCoeff}{0.00005}

\newcommand{\rraptorUnsignedMeanSqError}{0.001}
\newcommand{\rraptorUnsignedRTwo}{0.102}
\newcommand{\rraptorUnsignedIntercept}{0.206}
\newcommand{\rraptorUnsignedCoeff}{-0.00025}

\newcommand{\roboSignedMeanSqError}{0.002}
\newcommand{\roboSignedRTwo}{0.089}
\newcommand{\roboSignedIntercept}{0.774}
\newcommand{\roboSignedCoeff}{0.00031}

\newcommand{\roboASignedMeanSqError}{0.004}
\newcommand{\roboASignedRTwo}{0.286}
\newcommand{\roboASignedIntercept}{0.453}
\newcommand{\roboASignedCoeff}{0.00096}

\newcommand{\roboBSignedMeanSqError}{0.001}
\newcommand{\roboBSignedRTwo}{0.076}
\newcommand{\roboBSignedIntercept}{0.153}
\newcommand{\roboBSignedCoeff}{-0.00023}

\newcommand{\roboCSignedMeanSqError}{0.002}
\newcommand{\roboCSignedRTwo}{0.129}
\newcommand{\roboCSignedIntercept}{0.168}
\newcommand{\roboCSignedCoeff}{-0.00042}

\newcommand{\roboUnsignedMeanSqError}{0.002}
\newcommand{\roboUnsignedRTwo}{0.089}
\newcommand{\roboUnsignedIntercept}{0.226}
\newcommand{\roboUnsignedCoeff}{-0.00031}

\newcommand{\ftcDNCMeanSqError}{9.6e7}
\newcommand{\ftcDNCRTwo}{0.005}
\newcommand{\ftcDNCIntercept}{9,954.39}
\newcommand{\ftcDNCCoeff}{-1.5}

\newcommand{\ftcTotalComplaints}{13,707,524}
\newcommand{\ftcDNCStartTime}{2020-02-14}
\newcommand{\ftcDNCEndTime}{2024-05-23}
\newcommand{\ftcDNCDays}{1,560}

\newcommand{\pponeLabelSilhouetteCosine}{-0.38}
\newcommand{\pponeLabelSilhouetteEuclidean}{-0.17}
\newcommand{\pponeLabelCHScore}{17.1}

\newcommand{\roboSSA}{30.4\%}
\newcommand{\roboSSB}{21.6\%}
\newcommand{\roboSSC}{9.5\%}
\newcommand{\roboSSNone}{38.5\%}

\newcommand{\roboSSAOne}{51.2\%}
\newcommand{\roboSSBOne}{13.8\%}
\newcommand{\roboSSCOne}{14.2\%}
\newcommand{\roboSSNoneOne}{20.7\%}\clearpage{}
\clearpage{}\newcommand{\quotes}[1]{\textit{``#1''}}\clearpage{}

\title{Characterizing Robocalls with Multiple Vantage Points}

\author{\IEEEauthorblockN{Sathvik Prasad, Aleksandr Nahapetyan, Bradley Reaves}
\IEEEauthorblockA{\{snprasad,anahape,bgreaves\}@ncsu.edu\\
North Carolina State University}
}

\maketitle

\begin{abstract}
Telephone spam has been among the highest network security concerns for users
for many years. 
In response, industry and government have deployed new technologies and
regulations to curb the problem, and academic and industry researchers have provided 
methods and measurements to characterize robocalls. 
Have these efforts borne fruit?
Are the research characterizations reliable, and have the prevention and
deterrence mechanisms succeeded?

In this paper, we address these questions through analysis of
data from several independently-operated vantage points,
ranging from industry and academic voice honeypots to 
public enforcement and consumer complaints, some with over 5 years of historic
data.
We first describe how we address the non-trivial methodological challenges
of comparing disparate data sources, including comparing audio and transcripts
from about 3 Million voice calls. 
We also detail the substantial coherency of these diverse perspectives,
which dramatically strengthens the evidence for the conclusions we draw
about robocall characterization and mitigation while highlighting advantages
of each approach. 
Among our many findings, we find that unsolicited calls are
in slow decline, though complaints and call volumes remain high. 
We also find that robocallers have managed to adapt to
STIR/SHAKEN, a mandatory call authentication scheme.
In total, our findings highlight the most promising directions for future
efforts to characterize and stop telephone spam. 

\end{abstract}
 
\section{Introduction}\label{intro}

Robocalls have devastated consumer trust in the telephone network in the
United States, and the problem is growing in other countries. Some
estimates have claimed that as many as half of all calls are unwanted
spam or fraud~\cite{zdnetOverHalf}. While many consumers simply do not answer calls from
unknown numbers, those who do often hear misleading sales offers or
impersonations of government agencies or well-known brands attempting to
shake down their victims. These calls are effective and extract
millions of dollars each year from victims.

The onslaught of fraud and abuse has made countering robocalls a major
public policy issue in the United States. Presidential administrations, FCC and FTC
Commissioners, and state attorneys general are among those who have made
phone spam a key issue in their platforms. The attention has led to new
laws like the TRACED Act~\cite{fccTRACED} and a swath of changes to regulations by the
FCC and FTC. These regulations have expanded the definition of illegal
calls and allowed or mandated that carriers implement technical measures
to reduce or prevent illegal calls. Consumer providers have also
responded by engaging and advertising their spam-call labeling services.
Researchers, motivated by personal frustration and public need, have
also expended significant energy to develop tools and techniques to
characterize and counter spam calls.

With the substantial public and private investment in combatting the
problem comes a need to assess our progress and re-evaluate what we
believe we know about illegal calling. Industry sources publish regular
reports on call trends, but using proprietary and opaque methodologies.
For example, call
labelling\footnote{E.g., applying ``Scam Likely'' to caller ID information}
must happen with limited context, and is therefore difficult or
impossible to evaluate accuracy. Academic research uses transparent
methods and evaluations, but comprise ``one-shot'' efforts describing
a single moment in time. Collectively, the technical community and the
public at large need a reassessment of the state of the network.

The goal of this paper is to aid researchers, regulators, industry
operators, and the public in understanding and assessing robocalling and
what effects, if any, the totality of anti-abuse efforts have had. To
provide the most relevant view to-date, we fuse data from previously
studied resources (FTC Consumer Complaints and an academic voice
honeypot) and newly available vantage points (data from FTC's Project
Point of No Entry investigations, IRS fraud threat intelligence feeds,
and a commercial voice honeypot). All of these individual sources have
value and limitations. By comparing and contrasting these sources, we
can get a more complete picture of robocalling. In the process, we can
directly investigate how operational choices affect the findings,
providing valuable insights for future researchers. Accordingly, this
paper makes the following methodological contributions:

\begin{itemize}
\item
  We describe and evaluate call audio clustering for campaign
  identification using audio embeddings. Unlike prior work
  \cite{pbmr20}, embeddings can be computed deterministically in
  parallel, are vastly more efficient to query at the scale of data we
  study here, and remain comparable in accuracy.
\item
  We show how to reliably and repeatably compare campaigns across
  vantage points.
\item
  We develop techniques to determine if a campaign is interactive or
  static, permitting us and future work to track the growth of
  interactive illegal calling over time. Interactive calls hide
  information from honeypots and are analogous to cloaking detection in
  web phishing, and we anticipate generative AI will accelerate this trend.
\end{itemize}

\noindent We also contribute the following results (among others) and lessons for
stakeholders:

\begin{itemize}
\item
  Longitudinal data indicates a secular decline in robocall volumes,
  though the magnitude of the decline may range from 25\%-50\%. This contradicts
  findings from a longitudinal robocall study from 2020~\cite{pbmr20}. We cannot attribute the decline to
  any single factor.[Sec~\ref{subsubsection:data_source_comparison}]
\item
  We present the first analysis of the adoption of
  STIR/SHAKEN call authentication protocol by robocalls. Notwithstanding the decline
  mentioned previously, we show that remaining robocallers have
  succeeded in in assimilating to the new regime.
  We further discuss
  reasons that explain why STIR/SHAKEN has failed to dramatically reduce
  robocall volumes and propose recommendations to address them.[Sec ~\ref{subsubsection:data_source_comparison} and Sec~\ref{sec:discussion}]
\item
  Implementing interactivity in a voice honeypot dramatically improves
  outcomes. We saw up to a 60\% increase in collected call audio content, and
  nearly an order of magnitude increase in callback numbers collected in an interactive honeypot.[Sec~\ref{subsubsection:honeypot_interactivity}]

\item
  We show that estimates from a 2018 study on caller ID blocking effectiveness~\cite{ppag18} still hold. However, they all but fail when one source of caller ID blocklist is applied to another vantage point.[Sec~\ref{subsubsec:same_source_blocking}]
\item
  We also show that manual call campaign labeling, whether by consumers
  or experts may be inconsistent. We recommend researchers and engineers to adopt our automated
  approach and release our source code to identify robocall campaigns from bulk robocall audio recordings using audio embeddings.[Sec~\ref{subsubsection:honeypot_interactivity}]

\item Using traceback data, we find that about 99\% of campaigns subject to enforcement actions through the Project Point of No Entry (PPoNE) investigation ceased to operate after the enforcement actions. The small fraction of campaigns that did not cease to operate do so by establishing resilient operations that spread their calls over as many as 10
different service providers.[Sec~\ref{subsubsection:campaign_overlap_analysis_takeaways}]

\item
  We estimate that  PPoNE investigated campaigns
  associated with 5.5\% of all robocalls observed in this study, though on average the
  investigation lagged the first observed call by \textbf{387 days}.[Sec~\ref{subsubsection:campaign_overlap_analysis_takeaways}]
\end{itemize}
 \section{Background and Related Work}\label{bg}

In the 1990's, declining voice call costs due to deregulation led to a
dramatic rise in telemarketing calls in the US. While the majority of
these calls were for legitimate goods and services, people found them
bothersome. In response, Congress empowered the Federal Trade Commission
(FTC) to take measures to reduce unwanted calls.

The first measure was to set and maintain a policy called the
Telemarketing Sales Rule (TSR), and the second was to establish the
National Do Not Call Registry (DNC). Among its many provisions,
prerecorded telemarketing calls are prohibited entirely unless the
caller has written consent from the consumer. All other telemarketing
calls are also prohibited without written consent if a telephone number
is listed in the DNC. The TSR also requires all telemarketing calls to
transmit correct, accurate Caller ID (CLI) information. The FTC and
other regulator are allowed to seek civil penalties of over \$50,000 for
each and every violation of the TSR.

While the stakes are high, the TSR and DNC have done little to prevent
an epidemic of automated sales or scam calls reaching virtually every
American on a regular basis. The reason for this policy failure is
partially technical and partially economic. The economic issue is that
robocalling is so profitable that far more call center operators engage in
unlawful activity than regulators and attorneys general have resources
to bring cases against them.

Robocalling is profitable and easy because deregulation led to the
North American Public Switched Telephone Network consisting of a
federation of over 1,000 telephone service providers, most of whom use
VoIP to transmit calls. All networks, whether using VoIP or legacy call
technologies, deploy gateways at their network edge to protect their
internal networks and handle media and signaling protocol conversions.
Most calls transit many networks between their origination and
termination networks. In the process, only the barest information ---
the caller's claimed ID --- is transmitted from network to network. Each
network maintains records of every call it originates, terminates, or
transits, but these records only know the previous ``hop'' in the call
path. The result is that when an illegal call is reported, the call
source has to be identified through a laborious process called
``traceback.''

Network operators, at the mandate of regulators and legislators, have
implemented two schemes to address these problems. The first is to
mandate a clearinghouse for traceback requests called the Industry
Traceback Group (ITG) to serve as a central point of contact between
traceback seekers and network operators. The FCC has mandated that all
providers respond to traceback requests within a 24 hour period. The ITG
has drastically improved traceback responsiveness, conducting over 300 traceback requests per month.

The second scheme was a requirement for all VoIP operators to support a
call authentication mechanism called STIR/SHAKEN. STIR/SHAKEN adds an
additional authentication header to call setup messages that is intended
to allow the provider at the receiving end to identify the originating
network and authenticate that the call information (especially Caller
ID) is accurate.
STIR/SHAKEN specifies three attestation levels.
The highest level of authentication is the `A' level attestation, where the originating provider is (i) responsible for originating the call into the network, (ii) has a direct relationship with the customer (call originator) and the ability to identify them if required, and (iii) is directly associated with the phone number used to originate the call.
`B' level attestation indicates that the provider meets the first two criteria as the `A' level, but the provider does not directly own the phone number used to originate the call.
`C' level attestation indicates that the provider has no relationship with the call originator.  
In 2020, the FCC mandated\cite{fccCombatingSpoofed} the adoption of STIR/SHAKEN as a call authentication framework across the North American Phone Network, pursuant to the TRACED Act\cite{fccTRACED}.The mandate imposed a deadline of June 30, 2021. However, the mandate included exception for smaller service providers and provisions for extensions when appropriate.

STIR/SHAKEN is analogous to DKIM in that it asks as a sort of
``postmark'' on a call request to indicate its source. It does not
actually authenticate the source of the call, nor does it offer mutual
authentication. It also does not provide mechanisms that are compatible
with legacy infrastructure, so calls that originate, terminate, or
transit non-VoIP networks irrecoverably lose this information. In
addition to fundamental limits on what it guarantees or how compatible
it is with the network, the short deployment timeframe led to a host of
failures and incompatibilities~\todocite{SIPNOC-Josh} so
severe that many legitimate calls could not be authenticated~\cite{sipforumIndustryTraceback}.

Contemporary and subsequent research efforts provided alternatives that
modified existing legacy protocols~\cite{tdza16,tdza17} or used
in-band~\cite{rbt16} or out-of-band\cite{rba+17,cww+21,du23} protocols to provide
cryptographic mutual call authentication. Other efforts have explored 
preventing spoofing~\cite{mxss14,mxss16}, fingerprinting call audio or data
features~\cite{bpa+10,sen14}, automating tracebacks~\cite{jagertraceback}, intercepting calls with voice
assistants~\cite{srf17,psp+23},
or detecting or preemptively blocking calls
based solely on metadata or calling
patterns~\cite{bap07,bsg+11,jjs+12,lxl+18,lrp+18,upla20,wbsw09}.
Finally, researchers have studied the vulnerability of humans to phone
scams~\cite{tdza19,hdc20} and how they could be helped~\cite{sbm+20,hgs+23}.

Most similar to our own work is data-driven characterization of robocalls, 
particularly using call complaint data or voice honeypots.
While prior to 2012 there was a research emphasis on VoIP spam and other forms
of abuse~\cite{keromytis2009survey}, the focus was on the impacts to VoIP
infrastructure directly exposed to the public Internet. 
The first documented discussion of a voice honeypot was a M3AAWG report in
2014~\cite{m3aawg}, followed by work covering the deployment and initial
results~\cite{gsba15} of the described honeypot. 
This first honeypot paper
explored geographic trends and overlap with consumer complaint data over 1.3M
calls collected in a 7-week period, though audio analysis was limited to
manual case studies.
Soon after, a paper explored collecting
voice and SMS using paid mobile SIM cards, finding that a small vantage point
of 8 SIM cards gave a small vantage point of roughly 600 calls in 7 months. 
Pandit et al. studied the potential of honeypot and complaint data to serve as 
an estimate of Caller ID blacklisting effectiveness, claiming up to a 55\% 
unwanted call reduction~\cite{ppag18}. 
Finally, in 2020 Prasad and colleagues presented initial results from audio
and metadata analysis of 11 months of honeypot data, including describing a
method for clustering related calls together based on audio and showing a
steady rate of unwanted calls~\cite{pbmr20}. 
Their followup work~\cite{SnorCall} applied natural language processing to call
audio to analyze call content. 
Among other efforts, they demonstrated extraction of callback numbers that
could be used to identify and prosecute scammers.

Compared with prior work, the first principal contribution of this paper
is to substantially improve and add to the growing body of robocall
analysis techniques. Our new campaign identification approach scales to the
volumes of historical and current robocalls, and we are the first to study 
STIR/SHAKEN and the impact of honeypot interaction. As for results,
we confirm or refute prior findings with much greater confidence than before. 
We overturn previous findings of blacklist effectiveness and long-term static
call volumes, while we confirm previous studies of low overlap in honeypot and 
consumer complaint data. We also show that robocalling is such a large problem
that two large honeypots can have high call volumes but low overlap.

 \section{Methodology}
\label{sec:methods}

Collecting, processing, and extracting insights from telephony abuse data is non-trivial. We develop specialized techniques to process robocall audio data at scale. In this section, we describe these methods, rationalize the design choices, evaluate their performance, and use these techniques to derive insights from a multitude of robocall-related data sources.

\subsection{Audio Aggregation}
\label{sec:audio_aggregation}

\begin{figure}[t]
	\centering
    \includegraphics[width=.9\columnwidth]{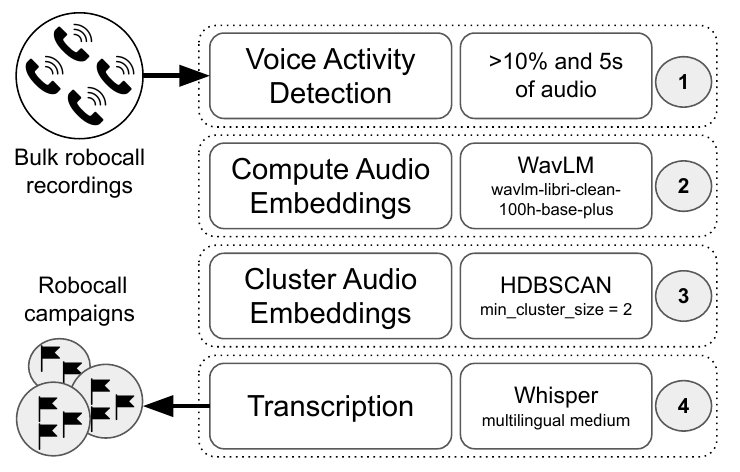}
    \caption{Audio processing pipeline}
	\label{fig:robocall-audio-processing-pipeline}
\end{figure}

A fundamental unit of measurement in the robocalling ecosystem is a \textbf{robocall campaign}. We define a campaign as a set of phone calls that play the same audio message. In the context of data collecting in honeypots, a campaign is a collection of audio recordings where the audio in each recording is identical or nearly identical. Our technique enables us to efficiently extract campaigns from bulk robocall audio data that spans millions of calls in a reasonable time with high accuracy on a commodity desktop machine with reasonable memory and CPU resources. A strict definition of a campaign allows us to rigorously evaluate the performance of our aggregation pipeline through manual analysis, due to lack of ground truth. Next, we describe each step of the audio aggregation pipeline shown in Figure~\ref{fig:robocall-audio-processing-pipeline}.

\subsubsection{Silence Detection}
\label{sec:silence_detection}

We use Voice Activity Detection (VAD) to identify and discard silent calls. Silent calls are common when characterizing unsolicited phone calls collected in a honeypot. Since they do not contain any audio content, they offer little value in the context of robocall campaign analysis. Using WebRTC VAD, we measure the duration of audio and the duration of silence for each call. Calls containing at least five seconds of audio and at least 10\% of audio content are retained for further processing. These thresholds are chosen based on estimates of how long a typical English speaker takes to deliver a meaningful short sentence in the context of a robocall campaign. In addition to receiving robocalls, honeypots receive unsolicited FAX calls and calls that merely play music-on-hold when answered. We identify and filter such calls at the transcription level by discarding calls with fewer than 30 characters in the transcript. After this pre-processing step, we discard silent calls and retain calls with meaningful audio content for further processing.

\subsubsection{Computing Audio Embeddings}
\label{sec:audio_embeddings}
We extract audio features from robocalls with substantial audio content by computing audio embeddings using pre-trained models. Recent advances in transformer-based speech language models enabled us to extract meaningful representations from audio data.
Using \textit{WavLM}~\cite{huggingfaceP}, we compute 768-dimensional embeddings from the audio of each robocall.
\textit{wavlm-libri-clean-100h-base-plus} model is a fine-tuned version \textit{WavLM-Base-Plus}.
The base model was pre-trained using more than 90,000 hours of human speech data and can capture human speaker characteristics.
The \textit{wavlm-libri-clean-100h-base-plus} model is the fine-tuned version of the base model for Automatic Speech Recognition (ASR) on the LibriSpeech dataset, which has a WER of 0.0683\cite{openslr}.
Since the model is pre-trained on 16kHz audio data, we resample robocall recordings to 16kHz before computing the embeddings.
By performing a forward pass of the audio data through the model, we extract the output of the hidden states.
The base model architecture has 12 Transformer encoder layers\cite{attention}, a hidden state of 768 dimensions, and 8 attention heads.
We aggregate the embeddings at the last hidden state using mean pooling to obtain a single 768-dimensional embedding for each audio recording.
Since the \textit{WavLM} model architecture captures human speaker characteristics, the computed embeddings represent the human speech within the robocall audio recordings.
\cite{WavLM}

\subsubsection{Advantages of Embeddings} 
The audio embedding-based approach offers several advantages over fingerprint-based audio analysis.
Audio fingerprint-based techniques require a database of fingerprints to compare against.
This leads to space and time complexity tradeoffs, as we need to compare hundreds of thousands of calls against each other.
The computational complexity and the memory footprint grows rapidly, making it infeasible to scale to the volume of data presented in this work.
In contrast, embeddings are compact vectors that can have a notion of similarity in the vector space.
This allows us to compare embeddings using an appropriate similarity metric, such as cosine similarity, which is not possible with fingerprints.
As a consequence, relative comparison drastically reduces the computational complexity and the memory footprint, making it feasible to scale to the volume of data presented in this work.
Furthermore, we can use distance-metric based cluster evaluation technique to measure the quality of robocall campaigns, whereas fingerprint-based techniques require manual evaluation.

\subsubsection{Clustering Audio Embeddings}
\label{sec:clustering}
After computing embeddings for each audio recording, we cluster the embeddings to identify robocall campaigns.
We use the HDBSCAN clustering algorithm to group similar embeddings into campaigns.
HDBSCAN is a density-based clustering algorithm that can cluster high-dimensional data.
Although most calls collected in honeypots are robocalls, there are mis-dials from real humans or calls from humans trying to reach the previous owner of the phone number.
Since these are not robocalls, our clustering approach must discard them as outliers.
Since HDBSCAN identifies outliers, it enables us to discard such misdialed calls and retain only robocalls for further analysis.
Finally, HDBSCAN is computationally efficient and converges relatively quickly when compared to other clustering algorithms\cite{hdbscanBenchmarkingPerformance}, making it a suitable choice for clustering the large volume of data presented here.

\subsubsection{Transcription and Language Identification}
\label{sec:transcription}

After identifying robocall campaigns, we transcribe the audio recordings within each cluster using Whisper's multi-lingual medium model.
Although we explored other STT models, we found that Whisper's speech-to-text models are one of the best-performing models~\cite{huggingfaceOpenLeaderboard} for transcribing noisy multi-lingual robocall recordings.
Another advantage of using Whisper to transcribe robocall recordings is that it is capable of identifying the language spoken in the audio.
The medium multi-lingual model~\cite{whisperHF} is trained on 100 languages and detects the language spoken in the audio by sampling the first 30 seconds of the audio.
During downstream analysis, we use the detected language to filter out calls that are not in English since our analysis is focused on English robocalls.

\subsection{Performance Evaluation of Campaign Detection}

Prior work has used audio fingerprinting techniques to identify and cluster robocall campaigns. However, such techniques fail to scale to the volume of data presented in this work, where we process data that is at least an order of magnitude larger than previously reported datasets of robocall audio recordings. 

\begin{table}[h]
    \centering
    \resizebox{0.48\textwidth}{!}{
\begin{tabular}{@{}l>{\centering\arraybackslash}>{\centering\arraybackslash}p{0.4in}>{\centering\arraybackslash}p{0.4in}>{\centering\arraybackslash}p{0.45in}>{\centering\arraybackslash}p{0.5in}>{\centering\arraybackslash}p{0.65in}>{\centering\arraybackslash}p{0.6in}@{}}

        \toprule
        Method    & {Total Clusters} & Max Cluster Size & {Total Calls Clustered} & {Data \newline Clustered \newline (\%)} & {Cluster Perfection \newline (\%)} & {Intra Cluster Precision \newline (\%)} \\ \midrule
    Wav2Vec2  & 119           & 44             & 740            & 67.27             & 94.11             & 97.68                 \\
    WavLM     & 145           & 51             & 842            & 76.55             & 93.79             & 97.84                 \\
    Echoprint & 165           & 19             & 586            & 53.27             & 96.97             & 98.83                 \\ 
    \bottomrule
    \end{tabular}
    }
    \caption{Campaign Identification Performance Evaluation}
    \label{table:performance_evaluation_tab}
\end{table} 

We evaluate the performance of our audio aggregation and clustering techniques using a public real-world dataset of robocalls collected from various sources~\cite{robocallDatasetTechReport}. We benchmark them against the fingerprint-based technique used in prior work~\cite{SnorCall,pbmr20}, and compare the performance of \textit{WavLM} and \textit{Wav2Vec2} based embeddings.
We find that \textit{WavLM} audio embedding-based approach outperforms the others while scaling to the volume of data presented in this work.
\textit{WavLM} embeddings cluster the highest fraction of calls into campaigns (76.6\%), while maintaining a high cluster perfection score (93.8\%) and intra-cluster precision (97.8\%).
It is also able to uncover large campaigns while discarding individual examples of robocalls that are not part of a campaign.

Prior fingerprint approaches~\cite{pbmr20} used ad-hoc metrics, described in Figure~\ref{fig:clusterperfect} and Figure~\ref{fig:intracluster}.
By manually listening to the audio recordings and by reading the transcripts, we label each cluster. While computing cluster perfection score, we use a strict criteria where we mark a campaign as a bad campaign even if a single misplaced call is present in the audio cluster.
The Intra-cluster precision scores for Wav2Vec2, WavLM and fingerprint-based techniques are 97.7\%, 97.8\%, and 98.8\% respectively.
The Cluster Perfection scores for Wav2Vec2, WavLM and fingerprint-based techniques are 94.11\%, 93.79\%, 96.97\%, as listed in Table~\ref{table:performance_evaluation_tab}.

\begin{figure}[]
\label{fig:clustering_box}
\centering
\includegraphics[width=0.6\columnwidth]{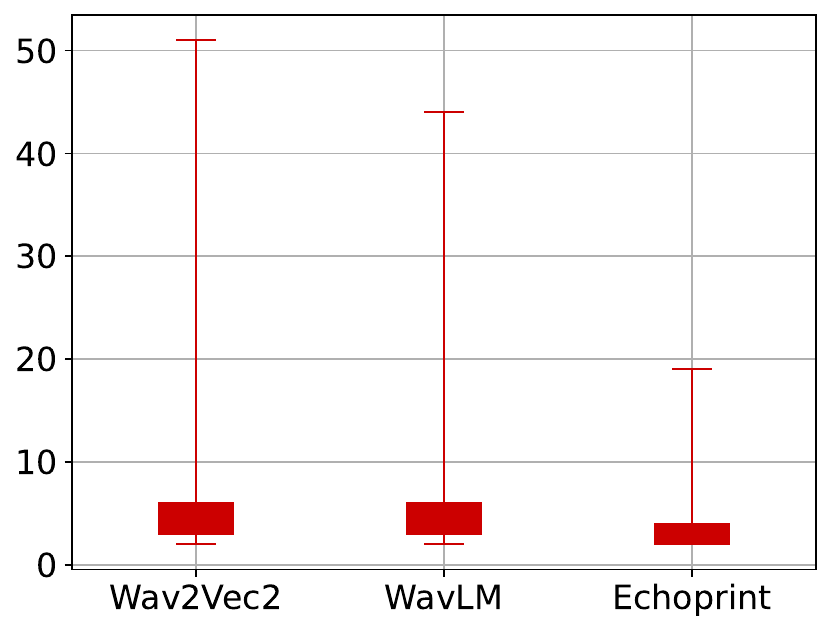}
\caption{WavLM and Wav2Vec2 uncover large (outlier) campaigns while maintaining a high median cluster size}
\label{fig:cluster_size_distribution}
\end{figure}

\myparagraph{Measuring campaign size} We analyze the distribution of cluster sizes and the size of outlier clusters to measure the ability of our techniques to uncover large campaigns, while maintaining good cluster quality. Figure~\ref{fig:cluster_size_distribution} shows the distribution of cluster sizes for each method, where WavLM and Wav2Vec2 maintain a higher mean cluster size compared to fingerprinting-based approach, while successfully identifying larger campaigns.

\myparagraph{Measuring data coverage} We evaluate the fraction of data clustered and the fraction of data discarded as outliers to measure the ability of our techniques to cluster robocalls into campaigns, while discarding individual examples of robocalls that are not part of a campaign and any misdials from human callers. 
WavLM based approach clusters the highest fraction of calls into campaigns (76.55\%).
Since it is able to uncover large campaigns while maintaining high cluster quality and performs at-par or better than two other techniques, we use WavLM for the rest of the analysis in this work.

\subsection{Comparing Campaigns Across Honeypots}
\label{sec:comparing_campaigns}

\begin{figure}[t]
	\centering
    \includegraphics[width=.9\columnwidth]{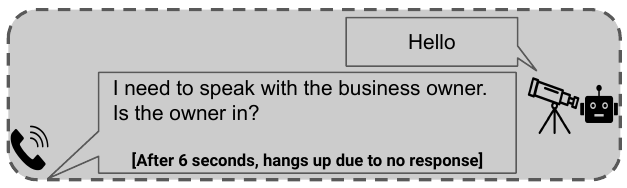}
    \caption{Interactive campaign disconnecting the call due to lack of engagement (no response) from \RoBo}
	\label{fig:interactive-call-with-robo}
\end{figure} \begin{figure}[t]
	\centering
    \includegraphics[width=.9\columnwidth]{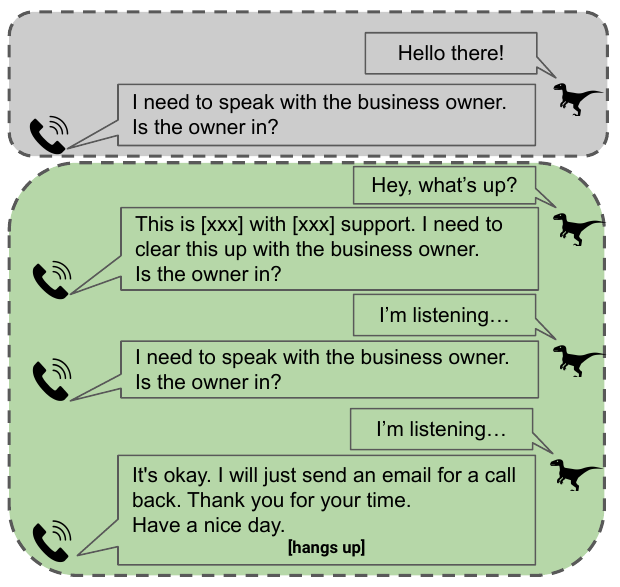}
    \caption{Interactive campaign engaging with \rraptor{} (an interactive honeypot) which yields additional details}
	\label{fig:interactive-call-with-rraptor}
\end{figure} 
After identifying robocall campaigns in English, we compare representative robocall transcripts of campaigns in each data source to identify common campaigns across multiple data sources.
We randomly sampled one representative transcript from each campaign and compared them against a randomly sampled representative transcript from campaigns uncovered in a different data source.
For each comparison, we compute a distance metric between the two transcripts to measure the similarity between the campaigns.
We build a similarity matrix by identifying the closest matching pairs.

\myparagraph{Comparing Transcripts vs Comparing Audio Embeddings} Using transcription to compute campaign overlap between two data sources is more reliable than audio embeddings for a few reasons.

\begin{itemize}
 \item \textit{Audio collection environment can differ}: Honeypots are often deployed with different configurations, lacking consistency in data collection. For example, different honeypots may set different maximum call durations, or use different codes (G711A, G729, etc) in their setup.
 
 \item \textit{Interactive vs Non-Interactive}: Comparing a two-channel audio stream from an interactive honeypot with a single-channel audio stream from a non-interactive honeypot is non-trivial. However, comparing the transcripts of audio coming from the calling party (robocaller) allows us to compare the content of the calls.
 
 \item \textit{Varying Network Conditions}: The network conditions under which the data is collected can substantially influence the audio quality (packet loss, jitter, etc.). A cloud-based vs. an on-prem honeypot also influences the network parameters of the raw audio. Transcripts are less sensitive to these variations.
\end{itemize}

\myparagraph{Transcript Similarity Metric}
Transcripts of phone call recordings are often noisy because of the underlying network characteristics.
The call audio is affected by packet loss, jitter, and other network conditions.
As a consequence, the transcripts from audio recordings that sound similar to a human ear may generate slightly different transcripts (different punctuations, capitalization, etc.).
Therefore, we choose a well-established string similarity metric.
We pre-process the transcripts into tokens\footnote[3]{SpaCy tokenizer using en\_core\_web\_sm model} by splitting them into words, removing numbers, punctuations, and special characters, and normalizing the case.

\myparagraph{Jaccard Similarity over Tokens} Jaccard similarity is a set-based similarity metric that measures the similarity between two sets.
Comparing tokenized transcripts with Jaccard similarity identifies campaigns that are identical or nearly identical in content.

\myparagraph{Longest Common Subsequence (LCS) over Tokens} LCS is a string similarity metric that measures the similarity between two strings by finding the longest subsequence that is common to both strings.
Using a similar tokenization and pre-processing approach, we use LCS over tokenized transcripts to capture campaigns that have the same initial pitch but behave differently after the initial greeting. 
Such campaigns are designed to wait for user engagement and continue their pitch only after they hear a response from the callee, as shown in the example conversation in Figure~\ref{fig:interactive-call-with-rraptor}.
Token-based LCS preserves the order of the tokens, and captures such interactive campaigns while also capturing non-interactive campaigns that have identical or nearly identical transcripts across different data feeds.

\myparagraph{Selecting a similarity threshold}
For all comparisons using LCS and Jaccard similarity, we use a similarity threshold of 90\% to determine common campaigns across multiple data sources.
A 90\% threshold allows us to capture campaigns accurately, while also accounting for minor variations in the transcripts due to network conditions, audio quality, and other factors.
The 10\% margin approximately translates to half a sentence (average transcript length is 70 words), which accounts for robocalls that terminate the call prematurely or may have minor transcription variations. 
With these thresholds, the expert author manually sampled and verified more than 100 example campaigns to ensure that these thresholds result in accurate matches across honeypots. 
By setting a high threshold, we prioritize precision over recall.

\subsection{Ringless Voicemail Detection}
\label{sec:ringless_voicemail_detection}
Ringless voicemail or voicemail injection campaigns are a type of robocall campaign where the audio message is delivered directly to the user's voicemail without ringing the user's phone.
These campaigns are noteworthy since they override the traditional call presentation behavior, where the user can see the incoming call along with crucial information like the "Spam/Scam Likely" label, the caller ID, call signer information, etc.~\footnote{In 2022 the FCC passed a rule which treats ringless voicemail as robocalls, and requires them to adhere to the TCPA: \url{https://www.fcc.gov/document/fcc-finds-ringless-voicemails-are-subject-robocalling-rules
}}
Voicemail injection campaigns bypass this call presentation and directly deliver the message to the user's voicemail.
A common approach to deliver ringless voicemail is to place an initial call so that the user's line is busy, and then place a well-timed robocall so that the call goes directly to voicemail.
We analyze the timing information in the call signaling information (SIP signaling) with matching calling and called numbers and measure the number of campaigns that adopt ringless voicemail delivery systems.

\subsection{Callback Number Extraction}
\label{sec:callback_number_extraction}

Callback numbers are phone numbers embedded within the call audio.
Robocall originator provides these numbers instructing the call recipient to engage with the call originator.
For legal telemarketing calls, callback numbers are toll-free numbers that allow the call recipient to opt-out of future calls.
Interestingly, some campaigns use low-quality text-to-speech (TTS) to read out the callback number, which makes it challenging to extract the callback number from the transcript.
For example, the callback number "844-924-XXXX" is transcribed as "eight four four nine two four...".
We refer to such callback numbers are \textit{vocalized callback numbers}.
We extract callback numbers from the transcripts of robocall campaigns using a curated set of regular expressions and use \textit{phonenumbers} library to process them further.

\subsection{Caller ID Normalization and Comparison}
\label{sec:caller_id_normalization}

We normalize caller IDs to E.164 format to discard invalid phone numbers and user entries in crowd-sourced data.
Since the size of our data spans tens of millions of entries, we implement an efficient tree-based caller ID search framework to compare caller IDs across data sources.
This framework ingests valid caller IDs from each feed, normalizes them to E.164 format, and stores them in a tree-based data structure, providing a logarithmic time complexity lookup API for caller ID comparison.

\subsection{Hardware Resources for Computation}
\label{sec:hardware_resources}

All the analysis presented in this work were performed using on-prem computing and storage resources.
Our workloads were distributed across two Intel i9 machines with four NVIDIA GeForce RTX 4090 in total in combination with a machine equipped with NVIDIA GeForce RTX 3070. It took multiple months of compute time and multiple TBs of storage to pre-process audio, compute audio embeddings and transcribe millions of robocall recordings. \section{Data Collection and Characterization}
\label{sec:data}

\begin{table}[ht]
    \centering
    \resizebox{0.48\textwidth}{!}{
        \begin{tabular}{@{}p{1.3in}|>{\centering\arraybackslash}p{0.45in}| >{\centering\arraybackslash}p{0.45in}| >{\centering\arraybackslash}p{0.275in}| >{\centering\arraybackslash}p{0.4in}@{}}
        \toprule
        {Vantage Point Name}               & Voice Honeypot                                                & Dataset Size & Call Audio & STIR/ SHAKEN \\ \midrule
        \RoBo                            & Yes                                                           & 10.05M      & Yes        & Yes         \\
        \rraptor                         & Yes                                                           & 948.76k          & Yes        & Yes         \\
        PPoNE Enforcement Actions        & No                                                            & 1.35k            & Yes        & No          \\
        FTC Do Not Call Complaints             & No                                                            & 13.71M        & No         & No          \\
        IRS APWG Data Feed               & No                                                            & 668.99k          & No         & No          \\ \bottomrule
    \end{tabular}
    }
    \caption {Vantage points, dataset sizes and characteristics}
    \label{table:data_sources}
\end{table} 
In this section, we describe each data source used in this paper, explain the data collection process, and discuss how we apply our data processing pipelines to each data source. We also discuss the ethical considerations of our work.

\subsection{\RoBo{}}
\label{subsubsection:robocall_observatory}

The \RoBo{} is a non-interactive automated call answering system designed to collect and store information about unsolicited phone calls.
It is a collection of VoIP phone numbers mapped to a PBX\todocite{asterisk}, where the PBX is configured to either answer or decline a phone call.
When answering a call, \RoBo{} plays a simple greeting --- ``Hello''.
It records any answered calls, and stored the audio as lossless WAV files.
It also stores the raw SIP signaling information (include STIR/SHAKEN headers) along with the call detailed records (CDRs).

\myparagraph{Data Charcterization} The data collected from the \Robo{} spans from \roboStartTime{} to \roboEndTime{}. During this period, the observatory received more than 10 million (\roboTotalCalls) calls. On average, there were approximately \roboCallsPerDay{} calls per day, with a high rate of calls per day during the initial years followed by a slow decline in the call rate over the years.

STIR/SHAKEN was enabled on the observatory on \roboStartSS{}. 
Since then, \RoBo{} has received \roboTotalCallsSinceSS{} calls.
Throughout the duration of \RoBos{} operation, about 30.40\% (\roboASS{}) of calls were signed with an A-level attestation, followed by 21.59\% (\roboBSS{}) with B-level attestation, 9.52\% (\roboCSS{}) with C-level attestation, and 38.48\% (\roboNoneSS{}) were unsigned calls.

The \Robo{} answered and recorded over 2.3 Million (\roboAnsweredCalls{}) unsolicited phone calls. 
Out of all calls recorded, only \roboPrecentValid{} (588,097) calls met our filtering criteria of containing enough audio information for further downstream audio processing. This is a decrease from the \whosCallingPrecentValid{} seen in prior work by Prasad\al{} \cite{pbmr20}.

The preprocessed calls containing substantial audio information were further processed to compute their respective audio embeddings using WavLM. These embeddings were clustered together using the campaign detection technique described previously in Section~\ref{sec:clustering}. A total of 360,723 (61.33\%) calls were clustered into 43,822 campaigns, with 227,374 discarded as outlier calls. These outlier calls are sometimes mis-dials or individual calls not part of a larger volume of traffic. The average campaign size was 8.23 calls per campaign, with some campaigns containing thousands of calls each. The largest campaign contained over 3,000 calls. The total time to cluster all calls was about 55 hours on a single machine with 32 cores and 128GB of RAM.

\myparagraph{Cluster Evaluation} To evaluate the quality of the clusters, we compute the silhouette score with cosine distance and the Calinski-Harabasz score, using sklearn's implementation. The clusters resulted in a silhouette score of 0.56 and a Calinski-Harabasz score of 1,166. Since the silhouette score ranges from -1 to +1, a score of 0.56 indicates that the clusters are well separated\cite{silhouette}. Additionally, a higher value of Calinski-Harabasz indicates dense and well separated clusters\cite{calinski1974dendrite}. Crucially, we were able to automatically compute these cluster evaluation metrics without any ground truth labels (unlike audio fingerprint based clustering), since each audio embedding can be mapped to a spatial domain.

\myparagraph{Language Distribution} Using OpenAI's Whisper, we found that 94.8\% of all calls clustered into campaigns were in English. The next prominent language was Spanish (2.5\%) and Chinese (1.5\%). The top three distribution aligns with the spoken languages in the United State, where English, Spanish and Mandarin are the top three spoken languages\cite{censusNearlyMillion}. The relative volume of Mandarin and Spanish robocalls in the \Robo{} were substantially higher in 2020-2021, and declined in 2022. Despite this decline, we uncovered numerous Spanish and Mandarin campaigns active in 2023-2024.

\myparagraph{Callback Numbers} Throughout \RoBos{} operation, we observed 6,010 callback numbers from 17,447 campaigns. About 3.7\% (222) callback numbers were vocalized\todowrite{What is this?} callback numbers, which were not captured by the techniques described by \cite{SnorCall}. Callback numbers continue to be an effective approach used by robocallers and telemarketers to engage with the target. In 2024 alone, we uncovered 247 callback numbers from 356 campaigns.

Out of the 6,010 callback numbers uncovered, 3,114 had a lifetime of less than 1 day, indicating that these callback numbers were used for short-lived campaigns.
Often, fraudulent or illegal campaigns use such short-lived callback numbers to evade detection.

The \Robo{} encountered 481 callback numbers with a lifetime of more than 1 year, with the oldest callback number that has been in use for over 5 years.
Often such callback numbers are used for useful robocalls and for legal telemarketing campaigns.
For example, callback numbers with lifetimes over many years are used by campaigns for public announcements (school closures, weather alerts, etc.), pharmacy notifications, and other legal telemarketing campaigns.
Many such legal campaigns provide a callback number, and instruct the target to call back to and ask the call originator to opt out of such calls from the campaign, similar to an unsubscribe mechanism in email marketing.\todowrite{Reaves wants to know if this is conjecture... I think so -Alex}

 \subsection{\rraptor{}}
\label{subsubsection:rraptor}

\rraptor{}~\cite{legalcallsonlyWhatRRAPTOR} is an interactive commercial honeypot that automatically answers calls made to the numbers assigned to it.
Unlike the \RoBo{}, \rraptor{} plays a set of pre-recorded prompts at various stages of a call to engage the caller in conversation.
It collects and stores audio data as two-channel mp3 files, and keeps track of the calling number, the date of the call and the STIR/SHAKEN attestation level of the call.

\myparagraph{Data Characterization} The data collected from \rraptor{}, spans from \rraptorStartTime{} to \rraptorEndTime{}. During this period, the honeypot received \rraptorTotalCalls{} calls at a rate of about \rraptorCallsPerDay{} calls per day, much higher rate than \RoBo{}.

STIR/SHAKEN was enabled throughout the duration of \rraptors{} operation.
About \rraptorSSAttestationA{} (\rraptorSSAttestationACount{}) of unsolicited calls received by \rraptor{} were signed with an A-level attestation, followed by \rraptorSSAttestationC{} (\rraptorSSAttestationCCount{}) with C-level attestation, \rraptorSSAttestationB{} (\rraptorSSAttestationBCount{}) with B-level attestation. About \rraptorSSAttestationNone{} (\rraptorSSAttestationNoneCount{}) calls were unsigned calls.

Unlike the robocall observatory, each call received by \rraptor{} was answered and recorded. Furthermore, the maximum audio duration of each recorded call was \rraptorMaxAudioDuration{}.
In total, \rraptorMoreThanTenpercentAndMoreThanFivesecPercentage{} (\rraptorMoreThanTenpercentAndMoreThanFivesec{}) calls met the criteria of containing more than 10\% of audio and more than 5 seconds of audio and were processed further to extract robocall campaigns. This is a slight decrease from prior work of \whosCallingPrecentValid{} and slightly higher then \RoBos{} \roboPrecentValid{}.\cite{pbmr20}

The robocall campaign aggregation process clustered \rraptorPercentageOfCallsClustered{} (\rraptorTotalCallsClustered{}) calls with substantial audio into \rraptorTotalCampaigns{} campaigns. The mean campaign size was \rraptorMeanCampaignSize{}, with the largest campaign containing \rraptorMaxCampaignSize{} calls. The total time to cluster the calls was about 20 hours on a machine with 32 cores and 128GB of RAM.

\myparagraph{Cluster Evaluation} The \rraptor{} clusters resulted in a cosine-distance based silhouette score of \rraptorSilhouetteScore{} and a Calinski-Harabasz score of \rraptorCalinskiScore{}. Both scores indicate that the clusters are well-separated and dense, similar to the clusters obtained from the robocall observatory data.\cite{calinski1974dendrite,silhouette}

\myparagraph{Language Distribution} Similar to the data collected in the robocall observatory, the majority of calls clustered into a robocall campaign in \rraptor{} were in English (\rraptorEnglishcallsPercetage{} or \rraptorEnglishcalls{} calls), with \rraptorSpanishcallsPercentage{} (\rraptorSpanishcalls{}) in Spanish and \rraptorMandarincallsPercentage{} (\rraptorMandarincalls) in Mandarin.

\myparagraph{Callback Numbers} Within a span of less than five months, \rraptor{} uncovered \rraptorTotalCallbackNumbersClustrered{} callback numbers, of which \rraptorCampaignsWithcallbackNumbersVocalized{} were vocalized\todowrite{What does thsi mean?}. These callback numbers were used across \rraptorCampaignsWithcallbackNumbers{} campaigns. \subsection{FTC Complaint Data}
\label{subsubsection:individual_feed_ftc_dnc}
The Federal Trade Commission's Do Not Call Complains feed consists of consumer reported complains of unwanted phone calls.
This data feed is available to the public via api.data.gov.
Each report consists of a mandatory field where the consumer reports the caller ID of the unwanted call and the date of the call.
It also contains a ``subject'' field where the consumer can select from 16 categories to describe the nature of the call.
The FTC's DNC feed spans from \ftcStartTime{} to \ftcEndTime{}, during which \ftcTotalReports{} reports were collected at a rate of about \ftcReportsPerDay{} reports per day. \subsection{PPoNE} \label{subsubsection:individual_feed_ftc_ppone}

As part of the Project Point of No Entry (PPoNE) initiative, the Federal Trade Commission (FTC) released a set of 31 enforcement actions against gateway service providers.
Each of these enforcement action contains a collection of traceback requests and their outcome, and is presented as a table at the end of the report.
We developed an automated system to download, process and extract information from these enforcement letters.
This automated pipeline parses the PDF enforcement letters and extract the audio URLs.
We also manually verified the results and used semi-automated PDF table parsing tools to extract tabular data embedded within the PDFs.

The \ppone{} dataset consists of a collection of \pponeTotalTracebacks{} audio examples ranging from \pponeStartTime{} to \pponeEndTime{}. 
Each audio example was used to complete a traceback request  to identify the source of the illegal traffic.
This is a small fraction of the total tracebacks conducted during this period, and is not representative of all tracebacks.

\begin{figure}[t]
	\centering
    \includegraphics[width=.985\columnwidth]{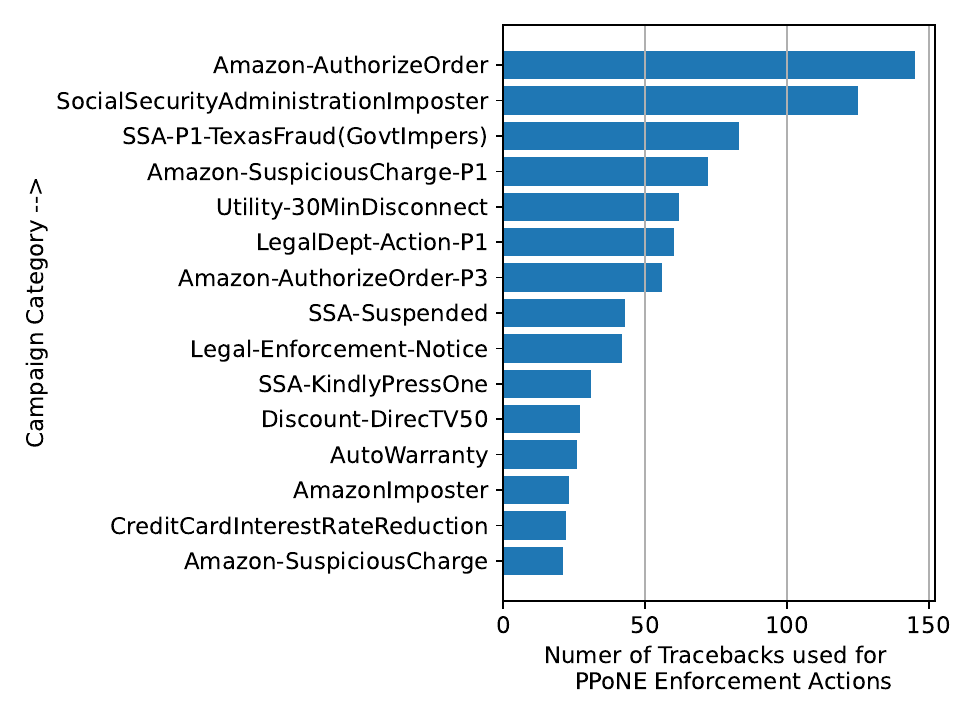}
\caption{Distribution of the top 15 categories of tracebacks of illegal robocalls used in PPoNE enforcement actions}
	\label{fig:ppone-analysis-plot-top-traceback-categories}
\end{figure}

This dataset crucially contains a collection of \pponeHumanLabels{} different label, where a label is assigned to each audio example, (eg: Amazon-AuthorizeOrder, SocialSecurityAdministrationImposter, AutoWarranty, etc).
Figure \ref{fig:ppone-analysis-plot-top-traceback-categories} shows the distribution of the top 15 labels in the dataset.
These labels are broad categorization of the type of the call based on the audio content.
These labels were likely assigned by human investigators, and are often quite broad --- including more than one campaign under the same label.

 \subsection{IRS APWG}
\label{subsubsection:individual_feed_irs_apwg}
The Anti-Phishing Working Group (APWG) IRS Impersonation Scam feed consists of a caller ID field that is supposedly associated with originating illegal IRS impersonation calls, or other fraudulent robocalls that are related to taxes or other tax related services.
The datafeed contains the caller ID and the timestamp of when the call was reported to the APWG community.
We use APWG's APIs to collect this data.
This feed was available to the members of the APWG community, but has since been discontinued.
The APWG\cite{APWG} data feed spans from \irsStartTime{} to \irsEndTime{}, during which \irsTotalCalls{} reports were collected at a rate of about \irsReportsPerDay{} reports per day. \subsection{Ethical Consideration}

The Institutional Review Board (IRB) and our university's legal team has reviewed and approved all of the data collection methods used in this paper. The maintainers of the commercial honeypot donated their data to the project. The \RoBo{} is set up in a first-party consent state, where only one party needs to consent to the recording of the call. When a call originates from outside the state, Federal law applies, which is also a first-party consent scenario. While we attempt to analyze automated mass call campaigns, we are conscious that the honeypots may receive calls from individuals. 
Through HDBSCAN discarding outliers and VAD-based pre-processing discarding short-duration calls, this work focusses on robocalling campaigns (collection of calls with substantial audio that are identical or nearly identical) and not individual call recordings.
The \RoBo{} data collection technique ensures that it gracefully terminates calls at 60 seconds (\rraptor{} at 90 seconds) to minimize data collection from non-robocalls. Furthermore, we have confidentiality agreements in place that prevent us from sharing raw data. 
For our other sources of threat intelligence, we have data sharing agreements with the APWG.
Finally, the PPoNE and the FTC DNC data were collected from public documents and APIs. \section{Results}
\label{sec:results}

This section characterizes the robocalling ecosystem from each independent vantage point, quantifies the relative overlap of caller IDs across multiple data feeds, measures the overlap of campaigns between \robo{} and \rraptor{}, quantifies the impact of PPoNE enforcement actions, and distills the key takeaways from these measurements.

\subsection{Metadata analysis}
\label{subsubsection:data_source_comparison}

\begin{figure}[t]
	\centering
    \includegraphics[width=.9\columnwidth]{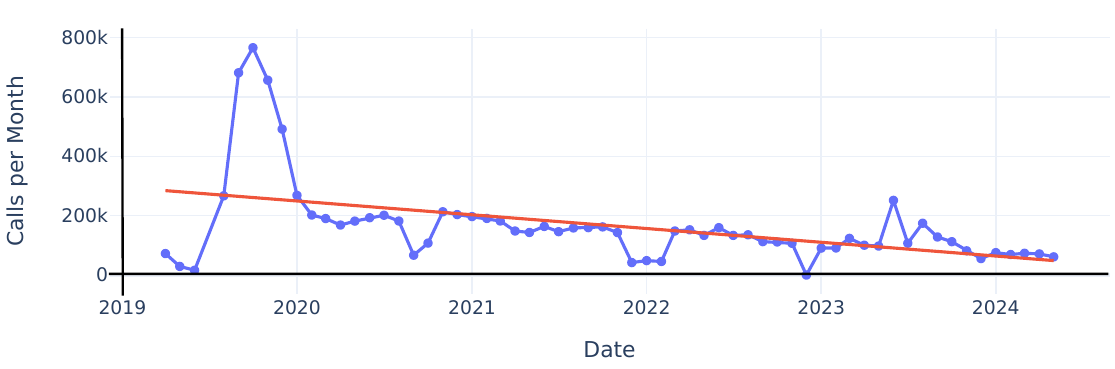}
\caption{\RoBos{} monthly call volume}
	\label{fig:honeypot-analysis-plot-monthly-call-volume}
\end{figure} 
\subsubsection{Robocall Volumes and STIR/SHAKEN Adoption}

One of the most visible changes to the telecom ecosystem in United States over the past few years is the mandatory deployment of STIR/SHAKEN.
We present the current state of the longitudinal robocall volumes and delve into how STIR/SHAKEN deployment has influenced the robocalling ecosystem from multiple vantage points.

\paperFinding{The volume of robocalls is declining in the long term
}
The monthly volume of robocalls received by \RoBo{} has been declining over the span of 5 years, as seen in Figure~\ref{fig:honeypot-analysis-plot-monthly-call-volume} in comparison to the flat rate seen in prior work\cite{pbmr20}.
A linear regression for the monthly volume of calls received by \RoBo{} had a slope of \monthlyHoneypotCoeff{} calls per month (\(R^2=\monthlyHoneypotR{}\)).
Although \RoBo{} has been receiving fewer calls over time, the reported sum of money lost by victims has increased from 2022 to 2023, as per the FTC's annual report\cite{ftcNationwideFraud}. A linear regression across the FTC-DNC daily complaints on the other hand, we see a slope of \ftcDNCCoeff{} complaints per day (\(R^2=\ftcDNCRTwo{}\)) over the span the 5 years.
Illegal robocalls continue to pose a significant threat to consumers in the United States.

\paperFinding{The volume of robocalls with STIR/SHAKEN attestation is increasing over the long term}
\RoBo{} started receiving STIR/SHAKEN attested calls from \roboStartSS{}.
Over a span of about 2.5 years, \RoBo{} observes an increasing trend of the fraction of calls signed using STIR/SHAKEN. In Nov 2021, about 60\% of all unsolicited phone calls were signed. Whereas, in May 2024, about 80\% of all calls were signed. Further, the volume of unsigned robocalls has been decreasing over time. In terms of overall traffic, using a linear regression we see the slope of unattested traffic to be \roboUnsignedCoeff{} normalized signed calls per day (\(R^2=\roboUnsignedRTwo{}\)).

\paperFinding{Attestation level of robocalls is increasing over time}
STIR/SHAKEN offers three levels of attestation --- A, B, and C, as described in Section~\ref{bg}.
As seen in Figure~\ref{fig:honeypot-analysis-plot-ss-all-signed-calls-2024-only} and Figure~\ref{fig:rraptor-analysis-plot-ss-all-signed-calls-2024-only}, majority of robocalls are signed with A-level attestation.
About \rraptorSSAttestationA{} (\rraptorSSAttestationACount{}) of unsolicited calls received by \rraptor{} were signed with an A-level attestation, followed by \rraptorSSAttestationC{} (\rraptorSSAttestationCCount{}) with C-level attestation, \rraptorSSAttestationB{} (\rraptorSSAttestationBCount{}) with B-level attestation. About \rraptorSSAttestationNone{} (\rraptorSSAttestationNoneCount{}) calls were unsigned calls. 
This trend is consistent with the data from \RoBo{}, where we see \roboSSAOne{}(179,544) of the calls signed with A-level attestation, followed by \roboSSBOne{}(49,937) with B-level attestation, and \roboSSCOne{}(48,442) with C-level attestation since 2024. The fraction of unsigned calls decreased from \roboSSNone{} in 2021 to \roboSSNoneOne{} in 2024, demonstrating a steady increase of STIR/SHAKEN signed robocalls.
This trend is consistent not only across our data feeds, but also with recent industry reports on robocalls signed with STIR/SHAKEN in 2024, according to which 80\% of robocalls are signed with A-level attestation~\cite{transnexusSTIRSHAKENStatistics}

\subsubsection{Role of Honeypots in Characterizing Robocalls}
\label{subsubsection:honeypot_interactivity}
\begin{figure}[ht]
	\centering
    \includegraphics[width=.83\columnwidth]{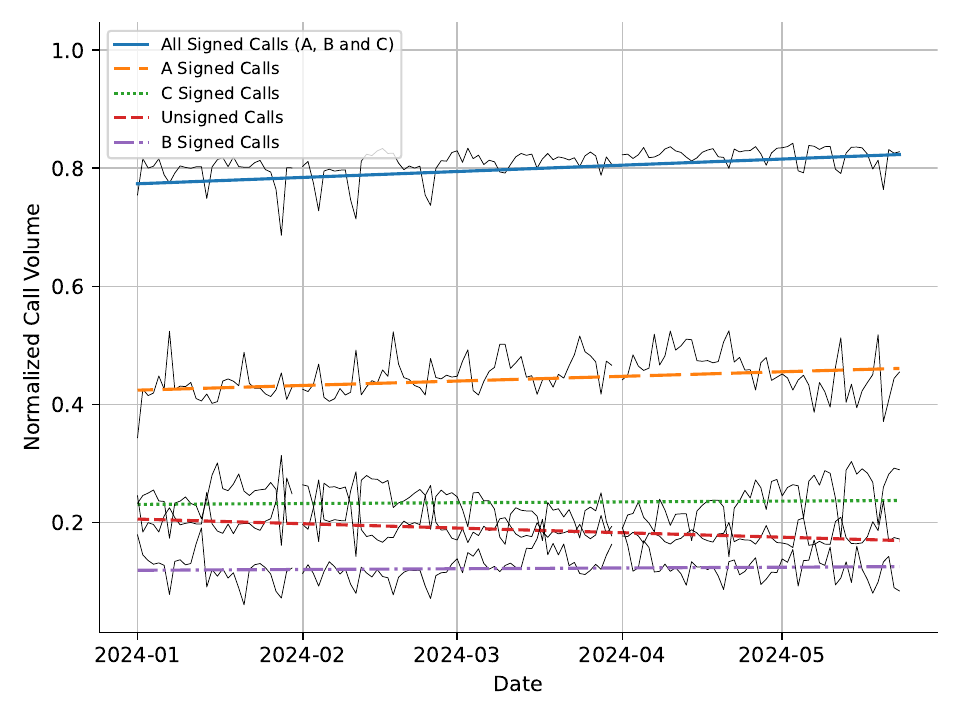}
\caption{Calls received by \rraptor{} signed with STIR/SHAKEN in 2024 (normalized by total calls per day)}
	\label{fig:rraptor-analysis-plot-ss-all-signed-calls-2024-only}
\end{figure} 
Honeypots serve as valuable vantage points to study robocalls.
We discuss and quantify the impact of adding interactivity to honeypots and how that can improve our visibility into the robocalling ecosystem.
 
\begin{figure}[ht]
	\centering
    \includegraphics[width=.83\columnwidth]{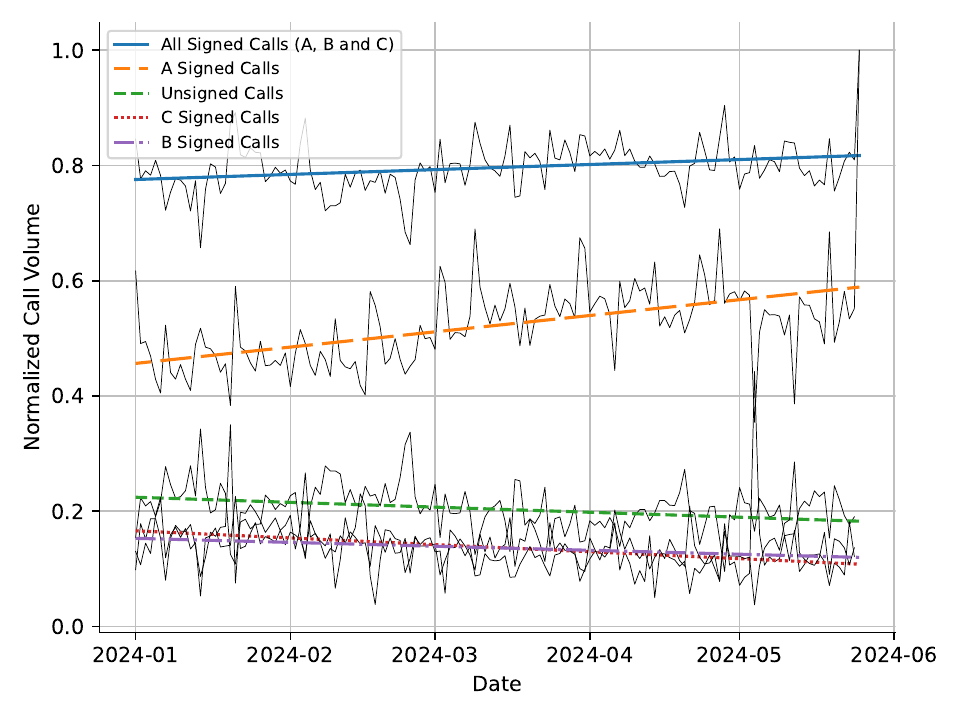}
\caption{Calls received by \RoBo{} signed with STIR/SHAKEN in 2024 (normalized by total daily calls)}
	\label{fig:honeypot-analysis-plot-ss-all-signed-calls-2024-only}
\end{figure}

\paperFinding{Large-scale and long-running honeypots are reliable vantage points to measure broader robocalling phenomena}
To study the extent to which honeypots can be used to measure broader phenomena within the robocalling landscape, we compare two phenomena inherent to the robocalling ecosystem.
Namely, fraction of unsolicited phone calls signed with STIR/SHAKEN and distribution of languages used by robocalling campaigns.
The fraction of unsolicited phone calls signed with STIR/SHAKEN is consistent across three distinct vantage points --- \rraptor{}, \RoBo{} and independent industry reports.
As seen in Figure~\ref{fig:rraptor-analysis-plot-ss-all-signed-calls-2024-only} and Figure~\ref{fig:honeypot-analysis-plot-ss-all-signed-calls-2024-only}, the fraction of signed calls is about 80\% across both honeypots as of May 2024, while following a similar trend throughout 2024.
Furthermore, A-attested calls are the most common, with C-attested calls being the second most while B-attested calls were the third most common category.
Finally, in both honeypots, the largest fraction of robocall campaigns use English.
Spanish and Mandarin are the second and third most common languages, respectively.
By comparing the signing rates and language distribution, we find that broader phenomena measured using \RoBo{} and \rraptor{} as two different vantage points are consistent.
This demonstrates that honeypots operating at the scale of \RoBo{} and \rraptor{} are reliable sources of data for measuring broader robocalling phenomena. They can independently report and provide valuable insights into the robocalling ecosystem.

Honeypots can differ in their operational characteristics by being more or less interactive, as described in Section~\ref{bg}\cite{m3aawg}.
However, developing, deploying and operating an interactive honeypot (versus a non-interactive honeypot) is substantially more challenging.
It requires well-timed interactive prompts, more computational resources, and non-trivial design overhaul to detect and play audio in near-real time.
Such fundamental changes to a honeypot's design is not justified if it yields only marginal benefits.

\paperFinding{
    An interactive honeypot captures more analysable audio from a robocall than a non-interactive honeypot}
We find that interactive honeypots are crucial to study the interactive campaigns that are active in the current robocalling ecosystem.
This is a notable shift from a previous study from 2020 by Prasad et. al.~\cite{pbmr20} where a non-interactive honeypot was sufficient to capture substantial amount of audio information.
\rraptor{} captured \rraptorMoreThanTenpercentPercentage{} calls with more than 10\% audio while \RoBo{} captured 42\% calls with more than 10\% audio.
Operationally, the maximum duration of recordings for \rraptor{} (interactive) was set to 90 seconds, while \RoBo{} (non-interactive) recorded calls were limited to 60 seconds.
Robocall originators are employing interactivity to limit the amount of information a honeypot can extract from a robocall.
This behavior is analogous to cloaking techniques used by phishing websites to evade detection.

\paperFinding{An interactive honeypot captures nearly 10 times more callback numbers from a robocall when compared to a non-interactive honeypot}
Information captured from a robocall extends beyond the amount of audio, such as callback numbers extracted from the audio.
In 2024, \rraptor{} captured \rraptorTotalCallbackNumbersClustrered{} unique callback numbers from \rraptorCampaignsWithcallbackNumbers{} campaigns, while \RoBo{} captured 247 callback numbers from 356 campaigns.
This may indicate that independent of the call volume the honeypot receives, employing an interactive honeypot to engage with the robocaller allows for more callback numbers to be extracted.
We discuss this phenomena in more detail in Section~\ref{subsection:campaign_overlap_analysis}.

\paperFinding{
    The lifetime of a callback number can vary significantly, with some being used for a short duration and other being used for extended periods of time 
}
To better understand the life cycle of callback numbers, we computed the lifetime of callback numbers by measuring the first and the last instance when a callback number was used by a campaign found in \RoBo{}. We see that the average lifetime of a callback number was 88 days, with a standard deviation of 219 days.

\paperFinding{
    We found that 362 campaigns were using voicemail injection techniques to deliver the message
}
Since \RoBo{} collects both signaling and call audio, this allows us to detect campaigns that leave a voicemail instead of a traditional phone call. We uncovered 3,784 campaigns with at least one call containing more than one call attempt within a 15 second window. We further filtered these campaigns to determine if more than 90\% of calls within the campaign had more than one call request message (SIP INVITE) within 15 seconds. 
A total of 362 campaigns were using voicemail injection techniques to deliver robocalls.
Such calls directly reach the user's voicemail and bypasses the standard call presentation to the user\cite{sbm+20}, where labels like ``Scam Likely'' and ``Spam Risk'' are displayed when the phone is ringing.

\paperFinding{
    Audio-based campaign identification is more specific than human-assigned labels
}
The PPoNE dataset described in Section~\ref{subsubsection:individual_feed_ftc_ppone} contains campaign names that are human interpretable, and were likely assigned by human analysts.
A total of \pponeHumanLabels{} such labels exist in the PPoNE dataset.
However, on clustering these raw audio recordings using the audio aggregation pipeline described in Section~\ref{sec:audio_aggregation}, \pponeReclustedTotalCampaigns{} campaigns were identified.
These clusters had a cosine distance based Silhouette score of \silhouetteScoreCosine{} and Calinski Harabasz Score of \calinskiHarabaszScore{}. Meanwhile, when clustered using the PPoNE human labels, the Silhouette score was \pponeLabelSilhouetteCosine{} and Calinski Harabasz Score was \pponeLabelCHScore{}.
The lack of consensus on the definition of a robocall campaign makes it challenging to reliably draw conclusions from human-assigned labels.
Therefore, using reliable and automated campaign identification techniques like the one presented in Figure~\ref{fig:robocall-audio-processing-pipeline} is crucial to track campaigns across data sources and study their evolution.

\subsection{Caller ID Overlap Analysis}
\label{subsubsection:callerid_overlap_analysis}

Each vantage point listed in Table~\ref{table:data_sources} contains caller ID information. Comparing the relative overlap in caller IDs contained within these feeds offer insights into effective strategies of harnessing originating phone number based threat intelligence about robocalling campaigns. To quantify the relative overlap between these data feeds, (i) we identify the temporal overlap between two feeds, (ii) measure the total number of reported ``events'' (calls, complaints, etc), (iii) identify unique originating phone numbers that are valid NANP numbers, and (iv) report the overlap between the two feeds.
Table~\ref{table:callerid_overlap_comparison} describes these fine-grained results.

\subsubsection{Honeypots and Crowd-sourced Consumer Complaints}
We compared the caller IDs uncovered by the two honeypots (\rraptor{} and \RoBo{}) with the FTC Do Not Call consumer complaints.
Over a span of about 4 months, 21,225 unique phone numbers found in \rraptor{} were also reported by frustrated consumers to the FTC via the Do Not Call complaints reporting mechanism. This overlap translates to about 4\% of the unique phone numbers in \rraptor{} and 8\% of unique phone numbers in the FTC DNC feed.
The top three categories consumers selected when reporting these calls were ``Others'', ``Dropped Call or No Message'' and ``No Subject Provided''.

Unlike \rraptor{}, \RoBo{} data collection overlapped with the FTC DNC for over 4 years.
During this time, \RoBo{} collected 6.7 million calls, and the FTC DNC received 12.9 million complaints, which is almost twice the number of calls received by \RoBo{}.
306,224 unique and valid NANP phone numbers uncovered by \RoBo{} were also reported by consumers to the FTC via the Do Not Call complaints.
This is the largest overlap in magnitude between any two feeds which translates to about 10\% of the unique phone numbers in \RoBo{} and 4.4\% in the FTC DNC feed.
The top three categories reported by consumers for the calls intersecting \RoBo{} and FTC DNC were ``Other'', ``Dropped Call or No Message'', and ``No Subject Provided''. This reinforces the observation that a large proportion of unsolicited phone calls are silent calls.

\subsubsection{\ppone{} Enforcement Actions and Caller ID Feeds}
The PPoNE is a small dataset of selected examples used in enforcement actions by the FTC. Therefore it does not represent a large-scale data collection effort like the other feeds.
By comparing historical PPoNE data with newer feeds, we can detect if campaigns subject to prior enforcement are still active.

A total of 75 numbers referenced in PPoNE enforcement actions were also reported by consumers via the FTC DNC reporting mechanism.
Due to the nature and size of the PPoNE dataset, drawing conclusions on the scale of overlap is not significant. It is noteworthy that about 6\% of phone numbers in the PPoNE were reported to the FTC DNC registry, and in 56\% of cases the FTC DNC feed reported the phone number first. Among these overlapping calls, the top category selected by consumers in the FTC DNC complaints was ``Calls pretending to be government, businesses, or family and friends''.
The PPoNE dataset and IRS APWG feed overlap for about 2 years, however only 11 unique phone numbers were present in both.

    \begin{table}[]
        \centering
        \resizebox{0.48\textwidth}{!}{
\begin{tabular}{@{}l>{\centering\arraybackslash}p{0.6in} >{\centering\arraybackslash}p{0.6in} >{\centering\arraybackslash}p{0.6in} >{\centering\arraybackslash}p{0.4in}@{}}
            \toprule
                             & {Total Call Attempts} & {Unique Calling Numbers} & {Overlap (Unique Numbers)} & {Overlap Period (days)} \\ \midrule
    
        \robo                & 344,344                                 & 151,937                                          & \multirow{2}{*}{15,692}                             & \multirow{2}{*}{143}                \\
        \rraptor             & 948,775                                 & 544,129                                          &                                                   &                                     \\ \midrule
    
        \robo                & 8,428,978                               & 3,143,974                                        & \multirow{2}{*}{9,190}                              & \multirow{2}{*}{1,415}             \\
        IRS APWG             & 668,991                                 & 262,931                                          &                                                   &                                     \\ \midrule
    
        \robo                & 6,712,525                               & 3,001,945                                        & \multirow{2}{*}{306,224}                            & \multirow{2}{*}{1,560}             \\
        FTC DNC              & 12,982,167                              & 6,885,768                                        &                                                   &                                     \\ \midrule
    
        \robo                & 4,356,546                               & 1,942,778                                        & \multirow{2}{*}{35}                                & \multirow{2}{*}{1,015}             \\
        PPoNE                & 1,350                                   & 1,350                                            &                                                   &                                     \\ \midrule
    
        \rraptor             & 948,775                                 & 544,129                                          & \multirow{2}{*}{21,225}                             & \multirow{2}{*}{143}               \\
        FTC DNC              & 1,349,281                               & 281,582                                          &                                                     &                                     \\ \midrule

        \rraptor             & 948,775                                 & 544,129                                          & \multirow{2}{*}{9}                                & \multirow{2}{*}{{0}}               \\
        PPoNE                & 1,350                                   & 1,350                                            &                                                   &                                     \\ \midrule
    
        FTC DNC              & 9,772,625                               & 5,764,999                                        & \multirow{2}{*}{33,749}                             & \multirow{2}{*}{1,096}             \\
        IRS APWG             & 111,053                                 & 92,281                                           &                                                   &                                     \\ \midrule
    
        FTC DNC              & 8,365,227                               & 4,695,484                                        & \multirow{2}{*}{75}                                & \multirow{2}{*}{1,015}             \\
        PPoNE                & 1,350                                   & 1,350                                            &                                                   &                                     \\ \midrule
    
        IRS APWG             & 30,734                                  & 26,039                                           & \multirow{2}{*}{11}                                & \multirow{2}{*}{1,015}             \\
        PPoNE                & 1,350                                   & 1,350                                            &                                                   &                                     \\ \bottomrule
        \end{tabular}
        }
        \caption {Caller ID overlap across data feeds}
        \label{table:callerid_overlap_comparison}
    \end{table}

\subsubsection{\ppone{} Enforcement Actions and Honeypots}
The PPoNE dataset and \RoBo{} overlap for over 2 years during which \RoBo{} received about 4.4 million calls (both answered and unanswered) from 1.9 million valid NANP phone numbers.
There were 35 unique caller IDs in \RoBo{} that had a matching caller ID in the PPoNE feed, totaling 103 calls made to \RoBo{}.

The PPoNE enforcement actions and \rraptor{} do not overlap since the enforcement actions pre-date \rraptors{} data collection time period.
However, we still uncovered 9 phone numbers listed in the PPoNE enforcement actions generating robocalls to \rraptor{}. 
\todowrite{Reference the next section}
While caller ID should be treated as a weak signal of campaign activity, we examine campaign audio content in the next section to understand if robocall campaigns that were subject to prior enforcement action remain active.

\subsubsection{\RoBo{} and \rraptor{} (independent honeypots)}
There were 15,692 unique phone numbers that were observed in both honeypots (\RoBo{} and \rraptor{}) over a span of about 5 months. 
About 3\% of caller IDs seen in \rraptor{} were also seen in \RoBo{}. However, over 10\% of the caller IDs seen in \RoBo{} were also seen in \rraptor{}. This indicates \rraptor{} has a broader view of the robocalling ecosystem than \RoBo{}.

\subsubsection{\RoBo{} and IRS APWG} A total of 9,190 distinct phone numbers were found in both \RoBo{} and IRS APWG, which made up 3\% of the unique phone numbers in the IRS APWG feed, but 0.3\% of the unique phone numbers in \RoBo{}. Although the overlap period spanned more than three years, there was low overlap between these two feeds.

\subsubsection{FTC DNC and IRS APWG} A total of 33,749 unique phone numbers were shared between the feeds. 
This translates to 37\% of the unique NANP phone numbers in the IRS APWG, but less then 1\% of the unique phone numbers in the FTC DNC.
About 90\% of the common phone numbers across these two feeds were fist found in the FTC DNC.
Since the IRS APWG feed is curated towards tax scams, IRS impersonation scams, and other tax-related scams, we explored the overlap between the category selected by consumers when reporting to the FTC DNC portal. 
The subject for over 40\% of the reports that matched in the FTC DNC were labeled as ``Calls pretending to be government, businesses, or family and friends''.

\subsubsection{Key findings and takeaways}
\label{subsubsec:same_source_blocking}
\ \ \ \ 

In a 2018 study by Pandit et. al.~\cite{ppag18}, researchers applied a sophisticated methodology to design and characterize a caller ID blocklist. While we did not fully replicate this methodology, we can present an upper bound on effectiveness of caller ID blocking using various data feeds. The intuition is that the best a blocklist source can do is to receive a call from a unique caller ID, which can then be used to block all future calls originating from the same caller ID. Therefore, the upper bound for a single vantage point call blocking approach is the number of calls that
matched a caller ID within the blocklist.

\paperFinding{Caller ID blocking is only effective when a blocklist is derived from a specific feed and applied to block calls made to the same feed}
A blocklist curated from FTC DNC feed is 21.1\% effective when applied to block calls made to \robo{}, and is 12.1\% effective when applied to block calls made to \rraptor{}.
This is an upper bound because it assumes that the FTC DNC observed the call before the call was made to either of the honeypots.
However, we found FTC DNC lags honeypots and often reported a caller ID after they were seen in honeypots.
A blocklist curated from \rraptor{} when applied to block calls made to \RoBo{} is only 16.3\% effective.
Similarly, a blocklist curated from \RoBo{} when applied to block calls made to \rraptor{} is 9.8\% effective.

For comparison, blocking calls from caller IDs seen within the same feed would result in a blockrate of 42.6\% for \RoBo{}, and 55.3\% for \rraptor{}. These findings are consistent with findings from the 2018 study~\cite{ppag18}, where a block list was created and used to block future unwanted calls made to the same set of phone numbers. This means that cross-source blocking of calls based on the caller ID is only marginally effective. 

\begin{figure}[t]
	\centering
    \includegraphics[width=.9\columnwidth]{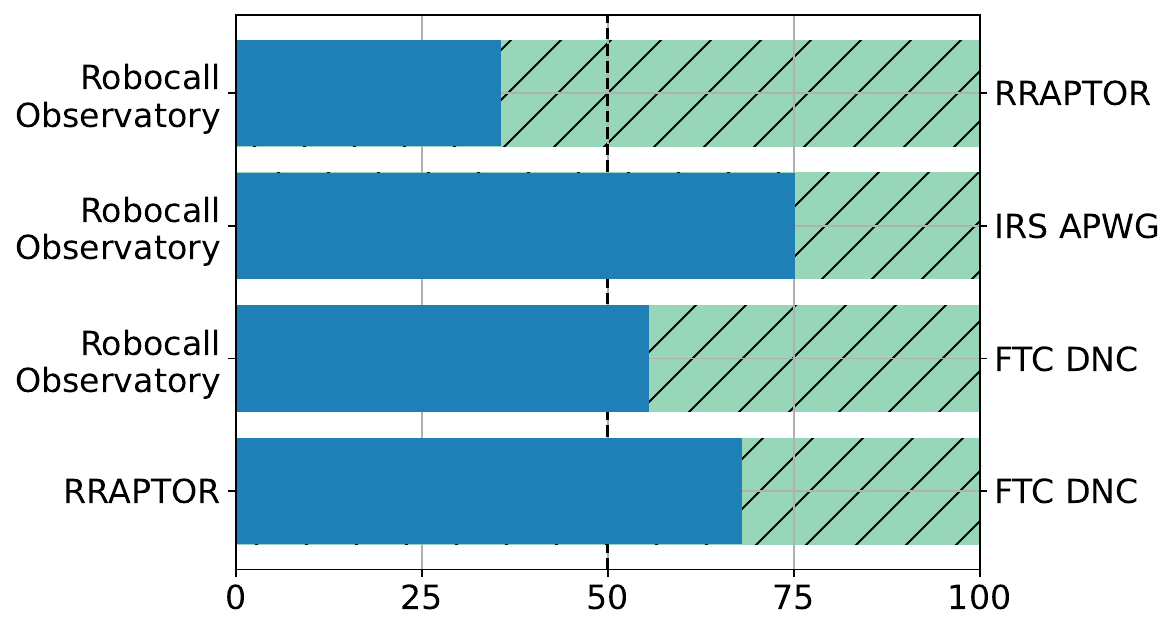}
    \caption{Honeypots consistently uncover the reported NANP phone before crowd-sourced feeds.}
	\label{fig:first_seen}
\end{figure} 
\paperFinding{Honeypots are able to capture active campaigns earlier than crowd-sourced consumer complaints}
Anecdotally, it is known that robocalling campaigns are observed first in honeypots before frustrating consumers to a point where they report these calls to the FTC's DNC or their respective service providers.
These reports often lead to tracebacks, and enforcement actions are taken by regulator agencies based on traceback results.
From Figure~\ref{fig:first_seen}, we see that in most of the cases honeypots observe the reported caller IDs even before they are reported by consumers.
Therefore, honeypots serve as early-visibility vantage points to track emerging robocall campaigns. 
\subsection{Campaign Overlap Analysis}
\label{subsection:campaign_overlap_analysis}

In this subsection, we analyze the overlap between the campaigns seen across \RoBo{}, \rraptor{}, and \ppone{} --- the three data sources that include call audio, as described in Table~\ref{table:data_sources}.

\subsubsection{Non-Interactive Campaigns seen in both \RoBo{} and \rraptor{}}
\label{subsubsection:campaigns_seen_in_both_robocall_observatory_and_rraptor
}

We compare representative transcripts of the campaigns seen in both \RoBo{} and \rraptor{} to identify the overlap between them, as described in Section~\ref{sec:comparing_campaigns}.
\RoBo{} and \rraptor{} overlapped for 143 days; for that timewindow, we process audio data to identify audio campaigns in \RoBo{} and \rraptor{}, independently.
We compare representative transcripts from each campaign in \RoBo{} and \rraptor{}, tokenize and pre-process the transcript, discard short transcript calls, compute the token-level Jaccard similarity and identify matching campaigns.
We find that 18.2\% of \rraptor{} clustered calls (6\% of campaigns) were also found in \RoBo{}; while from the vantage point of \RoBo{}, 16.6\% of clustered calls (14.25\% of campaigns) were seen in \rraptor{}.

By using the token-level Jaccard similarity as the distance metric, we ensure that calls and campaigns matched are identical or near-identical to each other.
Therefore, the overlap represents the campaigns that were seen by both the interactive and non-interactive honeypot, and behaved similarly in both honeypots.
This overlap represents the fraction of calls that do not adapt their behavior based on the honeypot's engagement level.

\subsubsection{Interactive and Non-Interactive campaigns seen in \rraptor{} and \RoBo{}}
\label{subsubsection:interactive_campaigns_seen_in_rraptor_and_robocall_observatory
}

We quantify the volume of interactive and non-interactive campaigns seen in \rraptor{} and \robo{} by following a similar approach, except that we use the token-level Longest Common Subsequence (LCS) similarity metric to compare the transcripts.
By using token-level LCS, we meet two requirements: (1) we ensure that identical calls are still flagged as identical, (2) we can now match calls that begin with the same script, but wait for a response from the honeypot in the middle of the call before engaging further (example in Figure~\ref{fig:interactive-call-with-rraptor}).

We find that 33.2\% of \rraptor{} calls (10.5\% of campaigns) were also found in \RoBo{}; while from the vantage point of \RoBo{}, 20.7\% of calls (19.24\% of campaigns) were also found in \rraptor{}. These overlap numbers represents the collection of interactive and non-interactive robocall campaigns seen in both \rraptor{} and \RoBo{}.
In both methods of identifying campaigns across feeds, the overlap between \rraptor{} and \RoBo{} is higher when looking at calls as part of campaigns, rather than caller-IDs.

\subsubsection{Campaigns seen in \RoBo{} that were subject to enforcement actions}
\label{subsubsection:campaigns_seen_in_robocall_observatory_that_were_subject_to_enforcement_actions}
To understand the extent to which the campaigns in \RoBo{} were subject to enforcement actions, we compare the campaigns seen in \RoBo{} with the calls in the PPoNE data.
Although the PPoNE data doesn't represent all enforcement actions, it does provides the first lower bound estimates of campaigns seen in \RoBo{} that were subject to enforcement action.
We process the calls in the PPoNE data to identify campaigns using our audio campaign detection pipeline, described in Section~\ref{sec:audio_aggregation}.
Each call in the PPoNE data represents a traceback.
Further, each call is a robocall (and not a misdial or a human call).
Therefore we do not discard the outliers and treat the outliers as separate campaigns.
Within a time period of over 2 years (1,015 days), we find that 5.5\% calls in \RoBo{} (19,735/360,723) were subject to enforcement action through the PPoNE initiative.
This represents 4.4\% (1,925/43,822) of campaigns seen in \RoBo{}.

\subsubsection{Campaigns that continue to operate despite enforcement actions}
\label{subsubsection:campaigns_that_continue_to_operate_despite_enforcement_actions}
The last reported call that was subject to enforcement action via the PPoNE initiative was on Oct 11th 2023.
However, a small fraction of illegal campaigns that were previously subject to enforcement actions continue to operate in 2024 and can be seen in commercial honeypots.
In \RoBo{}, we find that 0.5\% calls (53/9,548 clustered calls) and 0.49\% (8/1,621) campaigns active in 2024 were subject to enforcement actions via the PPoNE initiative.
In \rraptor{}, we find that 1.4\%  calls (1,448/103,947 clustered calls) and 0.62\% campaigns (94/15,084) active in 2024 were subject to enforcement actions via the PPoNE initiative.

\subsubsection{Takeaways from Campaign Overlap Analysis}
\label{subsubsection:campaign_overlap_analysis_takeaways}

\begin{figure}[]
	\centering
    \includegraphics[width=.9\columnwidth]{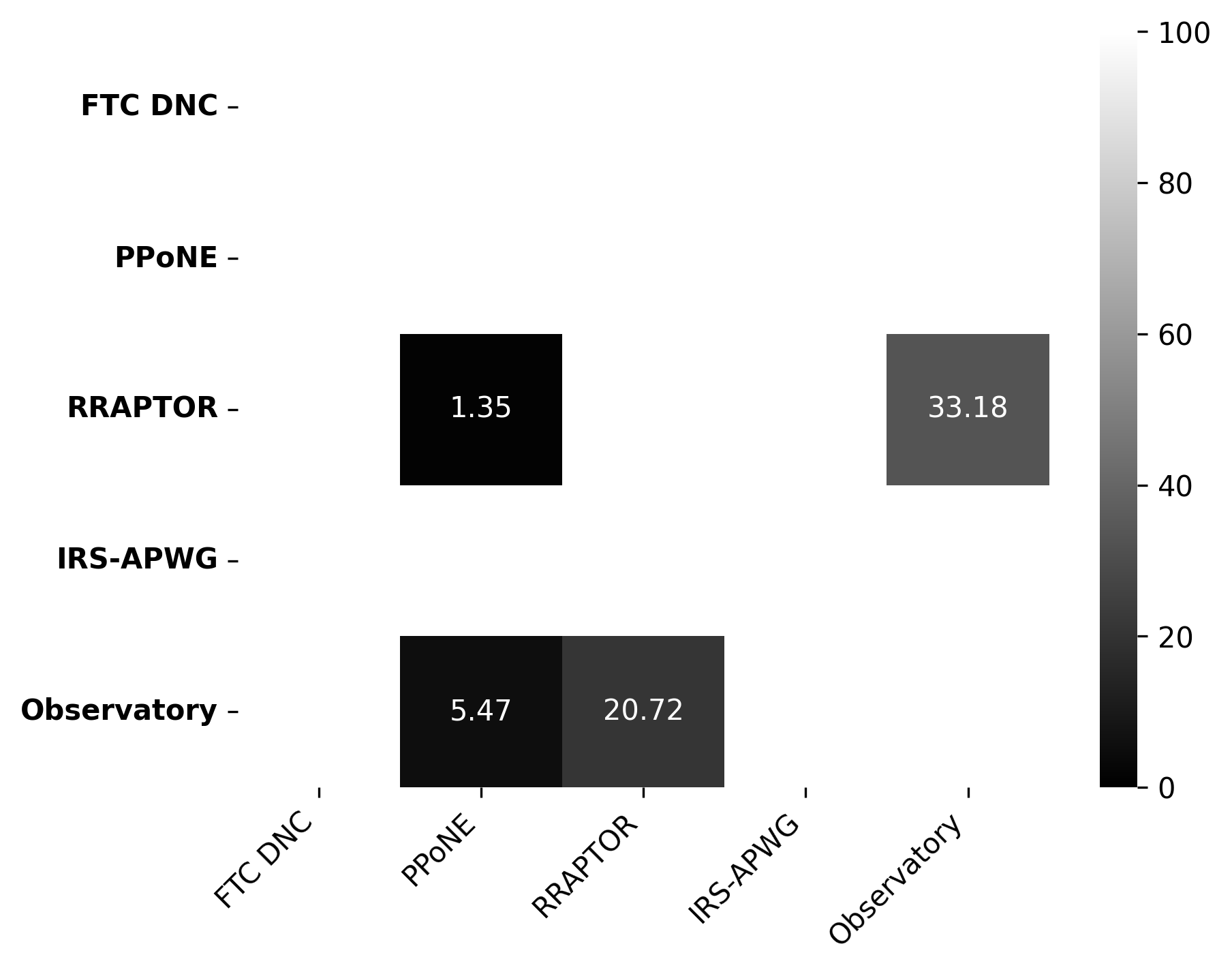}
    \caption{Heatmap of call overlaps of specific feeds relative to other feeds}
	\label{fig:call-overlap-heatmap}
\end{figure}

\begin{figure}[]
	\centering
    \includegraphics[width=.9\columnwidth]{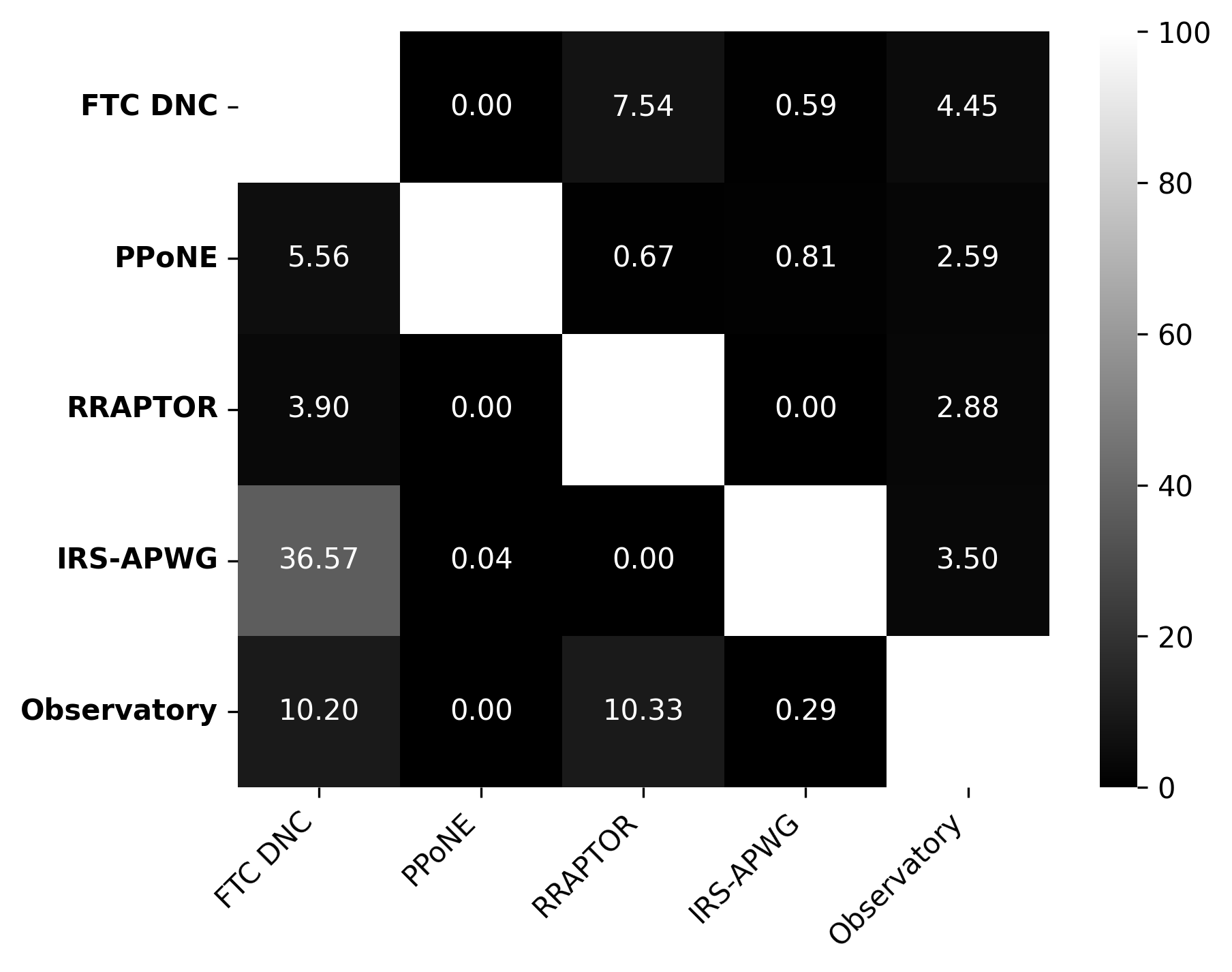}
    \caption{Heatmap of Caller ID Overlap Between Various Feeds}
	\label{fig:callerid-overlap-heatmap}
\end{figure}

\begin{figure}[]
	\centering
    \includegraphics[width=.9\columnwidth]{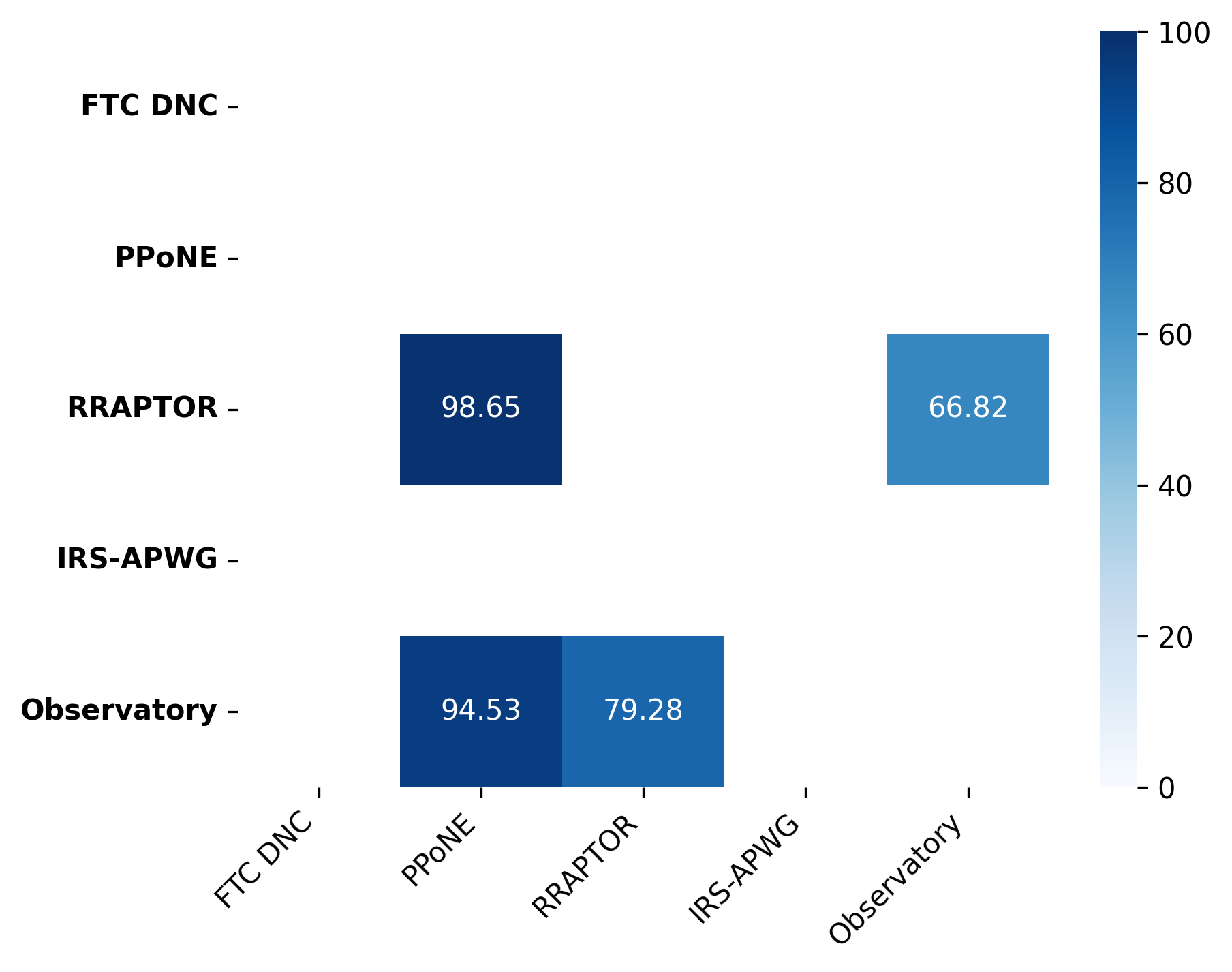}
    \caption{Heatmap of calls unique to specific feeds relative to other feeds}
	\label{fig:call-overlap-heatmap-neg}
\end{figure}

\paperFinding{Relying entirely on caller ID is an ineffective strategy to characterize illgal robocalling campaigns}
The relative overlap of caller ID across different data feeds are low, as seen in  Figure~\ref{fig:callerid-overlap-heatmap} and Figure~\ref{fig:call-overlap-heatmap}.
However, the relative overlap of calls based on campaign overlap analysis is much higher, as seen in Figure~\ref{fig:call-overlap-heatmap-neg} and Figure~\ref{fig:call-overlap-heatmap}.
This indicates that the fraction of activity captured by merely relying on caller ID is a much smaller fraction of total activity captured using campaign overlap analysis.
Therefore, caller ID is a useful feature to track illegal robocalls only when used in conjunction with other features or in tracebacks.
Importantly, caller IDs need to be understood as ephemeral and not as a persistent feature of robocalling campaigns.
\todowrite{Refernce the numbers from campaign overlap above}

\paperFinding{On average, \RoBo{} observed a call from a campaign that was subject to enforcement action about 387 days earlier}
We compute the time difference between the first call in a PPoNE campaign that was subject to enforcement action and the first call in the same campaign that was seen in \RoBo{}.
Out of a total of 101 PPoNE campaigns seen in \RoBo{}, 96 were seen first in \RoBo{}.
For these 96 campaigns, we find that the average time difference between the first call in a PPoNE campaign and the first call in the same campaign seen in \RoBo{} is 387 days.

\paperFinding{Illegal robocalling operations establish resilient operations by using multiple carriers to route their calls into the phone network}
In Section~\ref{subsubsection:campaigns_seen_in_robocall_observatory_that_were_subject_to_enforcement_actions}, we uncovered campaigns in \RoBo{} that were subject to enforcement actions.
The traceback data within the enforcement actions provides the details about the originating carrier that was responsible for routing the calls into the phone network.
Of the 101 PPoNE campaigns seen in \RoBo{}, we find that these campaigns used an average of 1.62 different carriers to route calls into the phone network.
Such operational strategies by illegal robocall originators enables them to continue operating despite the best efforts from enforcement agencies to block such calls by sending cease and desist orders to specific carriers using traceback data.

The most resilient campaign seen in \RoBo{} was distributing their traffic across 10 different carriers.
This campaign claimed to be from the Disconnection Department of an Electricity Company, warning the recipient about a disconnection of service within 30 minutes due to non-payment on the account.
The recipient was urged to press 1 to speak with a representative to avoid the disconnection.

The second most resilient campaign seen in \RoBo{} was using 5 different carriers to route calls into the phone network.
This campaign was impersonating Amazon and warning the recipient about a fictional order of Apple iPhone placed using their Amazon account, urging them to engage with the call to authorize or cancel the order.

\todowrite{Manually verified these claims with example transcribed in latex source}

\paperFinding{Illegal Robocalling Campaigns are persistent and continue to operate despite enforcement actions}
\todowrite{Refer to PPoNE matches in \rraptor{} or just move to the campaign overlap section}
In Section~\ref{subsubsection:campaigns_that_continue_to_operate_despite_enforcement_actions}, we found that most campaigns that were subject to enforcement actions via the PPoNE initiative stopped operating after the enforcement actions.
However, a small fraction of campaigns continue to operate despite enforcement actions.
By manually inspecting the active campaigns, we find that most of these active campaigns impersonate Amazon, and therefore are obvious scams.
The fact that these campaigns continue to operate despite enforcement actions suggest that these operations expect to be shut down by their respective carriers, and use alternate carriers to continue their operations.
Therefore, enforcement actions are crucial to shut down illegal operations.
However, we also need technical solutions that are swift and efficient in protecting consumers from illegal robocalls.

  \section{Discussion and Recommendations}
\label{sec:discussion}

\noindent \textit{Stakeholders need systematic and uniform campaign and topic
        classification}: 
        Outside of academic literature, the construct of a ``call campaign''
        is undefined. In academic usage, only virtually identical audio
        comprises a campaign. We found that human-labeled campaign identifiers in the PPoNE dataset
        mixed constructs. In some cases labelers mixed call
        recordings with thematically-similar but otherwise-different
        audio. We suspect that most fraud teams work with a
        similar mixed-construct model. Data
        sharing and collaboration is virtually impossible without a common 
        model of robocalling. By starting with a
        rigourous definition of campaigns, we were able to effectively
        compare campaigns across multiple data sources and modalities.

        Similarly, content classification is also ad-hoc. While the PPoNE
        labelers had free-text fields to write what seemed appropriate, 
        the FTC DNC complaint database is limited to those that were derived
        at the beginning of data collection. As the scam landscape has
        changed, so have our needs and what makes sense as an organizing
        criteria.  The top category consumers selected when reporting calls
        that we observed in \rraptor{} and \RoBo{} was ``Others''. More robust
        and updated categories can enable tracking of the types of scams
        active in the ecosystem from consumer complaints.

        Tracking illegal robocalls using caller ID alone has limitations, as
        discussed in Section~\ref{subsection:campaign_overlap_analysis} and
        Section~\ref{subsubsection:callerid_overlap_analysis}, but
        audio-based campaign comparison is more effective. The FTC DNC
        feed should enable consumers to upload audio recording examples or
        voicemail examples.
        Leveraging automated audio analysis techniques similar to the ones
        presented in this work can minimize manual analysis time and
        enable swift action against bad actors.

\noindent \textit{Stakeholders need tools to accelerate enforcement against telephone abuse}: 
        Enforcement actions  are key to combat illegal robocalls~\cite{fccIssuesFirst}.
        As seen in
        Section~\ref{subsubsection:campaigns_that_continue_to_operate_despite_enforcement_actions},
        enforcement actions ended virtually all relevant campaign activity.
        The challenge is that the technical and legal frameworks for doing so
        have high latency and low throughput.
        Investigators seem to be doing a good job prioritizing the top
        offenders, such as PPoNE targeting 5.5\% of all honeypot traffic
        with only 31 actions and 1,353 tracebacks. Despite incredible
        improvements to traceback, only a few hundred per month are feasible
        with current resources, and they still can take multiple days to trace
        a single call to its source. Moreover, the average campaign operated
        for \emph{over a year} before the investigation took place. 
        Stakeholders should look for ways to enable rapid enforcement based on
        early indicators available from honeypots. The recent 24-hour
        traceback requirements are a small yet promising step forward~\cite{fccFurtherExpands}.

        \textit{Robocall mitigation strategies should be driven by data-driven characterization.}
        The findings in Section~\ref{sec:results} demonstrate how the robocalling ecosystem adapts to changes intended to combat illegal robocalls. The deployment of STIR/SHAKEN, a call authentication framework intended to curb caller ID spoofing, has resulted in a growing trend of legitimate phone numbers being used to generate bulk robocalls that are authenticated by the originating carriers~\cite{itg_2024_report}.
        Carriers need to limit robocall originator's access to legitimate phone numbers and restrict their ability to generate illegal robocalls that are authenticated. To this end, developing and adhering to robust Know Your Customer and Know Your Upstream Provider policies and practices\todocite{} are crucial.

        \textit{Deployment of interactive honeypots to study ``smarter'' robocall campaigns.} Interactive robocalling campaigns continue to become more prevalent. 
        Easy-to-use generative-AI based audio services will likely accelerate this change.
        Widely deployed honeypots must adapt to this changing landscape. Researchers, anti-robocall product vendors, carriers and other stakeholders should develop better tools and techniques to study such interactive campaigns.

 \section{Acknowledgements}

The authors would like to thank the shepherd and anonymous reviewers for their helpful comments. 
We thank Bandwidth Inc. for providing VoIP service and phone numbers for the \RoBo{}.
We also thank the APWG, David Frankel and ZipDX for sharing the data that made this study possible.
This material is based upon work supported by the National Science Foundation under grant number CNS-2142930. This paper was partially supported by funds from the 2020 Facebook Internet Defense Prize.
We would like to thank Abhinaya for her helpful suggestions.
Any opinions,
findings, and conclusions or recommendations expressed in
this material are those of the authors and do not necessarily
reflect the views of the National Science Foundation, other funding agencies
or financial supporters. \bibliographystyle{IEEEtran}
\bibliography{bib/bibliography.bib,bib/relwork.bib}

\appendices
\label{sec:appendix}

\begin{figure}[t]
	\centering
    \includegraphics[width=.98\columnwidth]{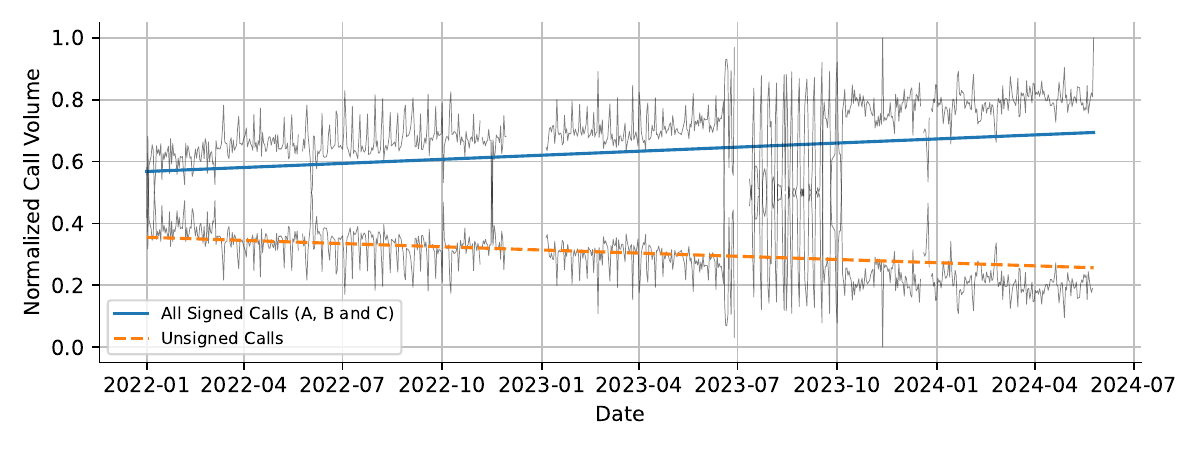}
\caption{Calls received by \RoBo{} Signed with STIR/SHAKEN from 2019 to 2024 (Normalized by total calls received per day)}
	\label{fig:honeypot-analysis-plot-ss-all-signed-calls-comprehensive}
\end{figure}   \begin{figure}[]
	\centering
    \includegraphics[width=.9\columnwidth]{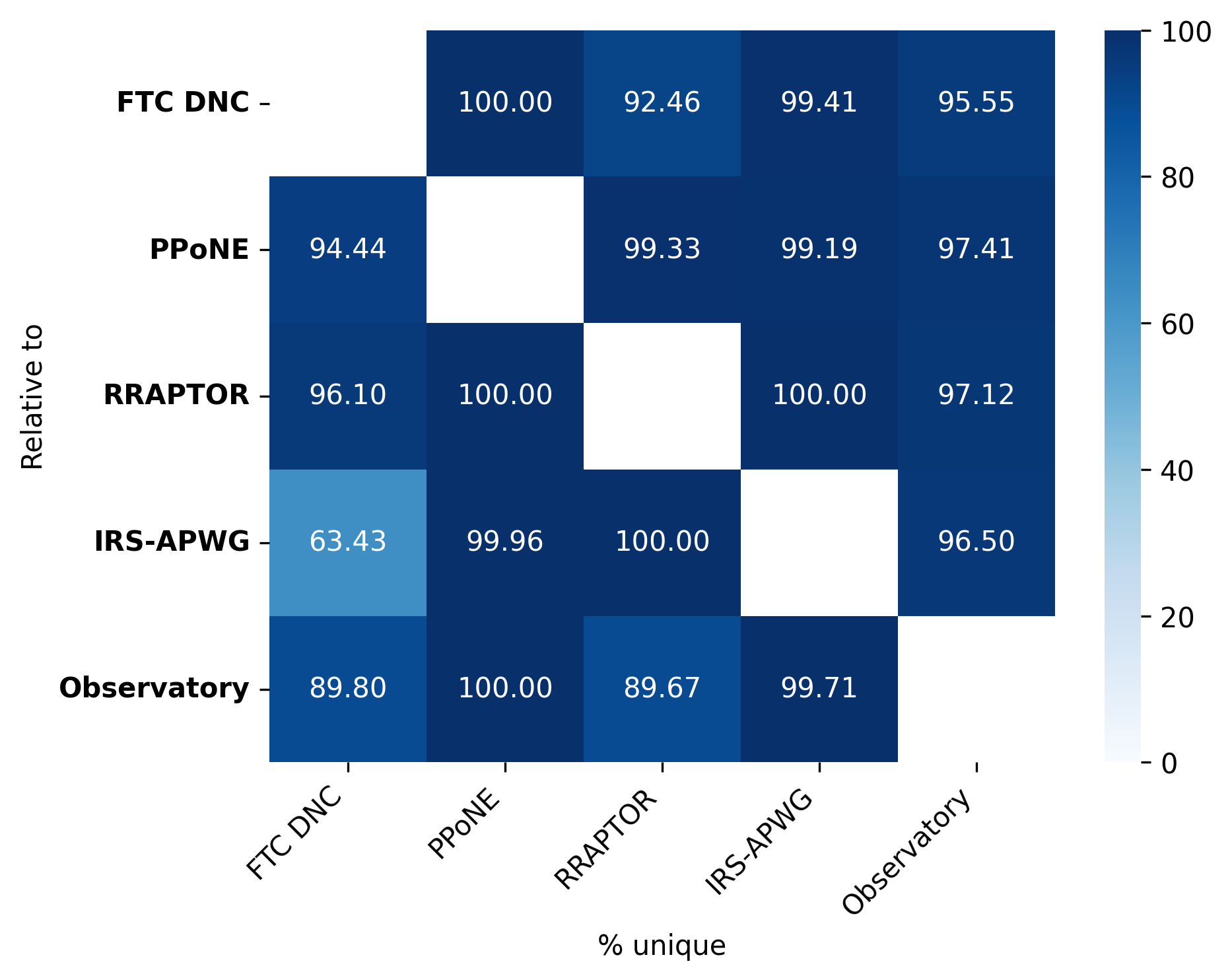}
    \caption{Heatmap of Caller ID Unique To Specific Feeds Relative to Other Feeds}
	\label{fig:callerid-overlap-heatmap-neg}
\end{figure}  \begin{figure}
   
   \begin{center}
         \(\left(\frac{\textnormal{Clusters without a misplaced call}}{\textnormal{Total clusters}} \right) \times 100\)
   \end{center}
   \caption{Equation used to calculate Cluster Perfection}
   \label{fig:clusterperfect}
\end{figure}

\begin{figure}
   
   \begin{center}
   \(\textnormal{Mean} \times \left( \frac{\textnormal{Correctly placed calls in the cluster}}{\textnormal{Calls in the cluster}} \right)\)
   \end{center}
   \caption{Equation used to calculate Intra-cluster Precision}
   \label{fig:intracluster}
\end{figure} %
 \clearpage{}
\newpage

\section{Meta-Review}

The following meta-review was prepared by the program committee for the 2025
IEEE Symposium on Security and Privacy (S\&P) as part of the review process as
detailed in the call for papers.

\subsection{Summary}
This paper presents a characterization of Robocall scams. The paper collects data from multiple vantage points and presents a longitudinal analysis. As part of the analysis, the authors cluster calls using audio embeddings, which allowed them to study robocall campaigns. As part of the characterization and analysis, the authors study overlaps between vantage points across different dimensions, like caller IDs. Overall this paper presents a longitudinal analysis of 3M voice calls over 5 years.

\subsection{Scientific Contributions}
\begin{itemize}
\item Provides a Valuable Step Forward in an Established Field
\item Addresses a Long-Known Issue
\end{itemize}

\subsection{Reasons for Acceptance}
\begin{enumerate}
\item This paper provides a valuable step forward in an established field by assisting researchers, regulators, industry operators, and the public in understanding the whole picture and the effectiveness of anti-robocall measures.

\item Addresses a Long-Known Issue. Understanding telephony scams is timely and highly important. Even though the paper shows that the amount of robocalls is going down, it is still a nuisance to anyone who owns a phone. This scourge does significant damage to trust in the telephony system, and therefore more insights into this area are important.

\end{enumerate}

\subsection{Noteworthy Concerns} \begin{enumerate} \item Definition of campaign. The paper defines a campaign as a "set of phone calls that play the same audio message." This notion of “campaign” is strict compared to other domains. A more versatile definition would consider calls with a common objective or from the same actor under the same campaign, even if the audio message differs. While it would be ideal to have a broader definition that accounts for same-source variance, unfortunately, there is no ground truth to evaluate against.

\item Having such a strict definition, the paper can guarantee that the call audio came from the same (audio) source — even without ground truth. However, the strict criteria may impact the findings obtained from the analysis of the campaigns and the scope of this limitation cannot be determined without properly evaluating the recall performance of the clustering.

\end{enumerate}

\end{document}